# Charge Symmetry Breaking and QCD[*]

Gerald A. Miller
Department of Physics, University of Washington, Seattle, WA 98195-1560, USA

Allena K. Opper
Department of Physics, The George Washington University, Washington, D.C. 20052, USA

Edward J. Stephenson
Indiana University Cyclotron Facility, Bloomington, IN 47408 USA

**Abstract**  Charge symmetry breaking (CSB) in the strong interaction occurs because of the difference between the masses of the up and down quarks. The use of effective field theories allows us to follow this influence of confined quarks in hadronic and nuclear systems. The progress in observing and understanding CSB is reviewed with particular attention to the recent successful observations of CSB in measurements involving the production of a single neutral pion and to the related theoretical progress.

CONTENTS

## 1 Introduction

Important insights into the properties of hadrons and nuclei can be gained by ignoring the often small effects created by the electric charge. For example, the proton and neutron have nearly identical masses and can be considered as two different states of a single particle, the nucleon. This leads to the expectation that the forces between nucleons should be independent of charge, once electromagnetic effects are removed. This charge independence of nuclear forces is also manifest in nuclear properties.

The isospin quantum number **T** was introduced so that the nucleon with isospin 1/2 has two states, in analogy to the spin quantum number. The third or $z$ component is associated with the charge, so that the charge of a nucleon is given by $e(1/2 + T_3)$ and that of an up or down (light) quark is $e(1/6 + T_3)$. Charge independence states that strong interaction ($S$) forces do not distinguish between the neutron and proton. As long as only the strong interaction is present, the isospin vector **T** can point in any direction so that the strong Hamiltonian obeys

$$[H_S, \mathbf{T}] = 0. \tag{1}$$

To satisfy this equation is to satisfy isospin invariance. The term "charge independence" is used often as a synonym, but truly applies only to the strong forces between nucleons.

The effects of the light quark mass difference and electromagnetism do not commute with the isospin operator. So isospin invariance and charge independence are only approximate symmetries. It is useful to distinguish between the two terms. For example, s-wave pion-pion scattering can be different between the $\pi^+ - \pi^0$ and $\pi^0 - \pi^0$ cases simply because the isospin Clebsch-Gordan coefficients give different combinations of the $T = 0$ and $T = 2$ strong amplitudes, making this scattering charge dependent even though the Hamiltonian obeys Eq. (1). Another example concerns the charge dependence and isospin breaking of the nucleon-nucleon (NN) $^1S_0$, isospin-1 scattering lengths. The matrix element of the non-vanishing operator $[H, T^2]$ between $T = 1$ states vanishes (so there is no isospin mixing), but the charge-dependent Coulomb interaction causes the scattering lengths of the $nn$ and $pp$ systems to differ.

The isospin rotation of 180° about the $y$ axis is the special transformation that defines the charge symmetry operator $P_{\mathrm{cs}}$:

$$P_{\mathrm{cs}} = e^{i\pi T_2}, \tag{2}$$

and charge symmetry holds provided

$$[H_S, P_{\mathrm{cs}}] = 0. \tag{3}$$

If charge symmetry is broken then isospin invariance and charge independence are broken, but the converse is not true. For isospin doublets one writes $\mathbf{T} = 1/2\boldsymbol{\tau}$ (where $\boldsymbol{\tau}$ are the three Pauli spinors). Then $P_{\rm cs}|u\rangle = -|d\rangle$ and $P_{\rm cs}|d\rangle = |u\rangle$. The term "charge symmetry" should not be confused or interchanged with the term "charge conjugation," which involves replacing a particle by its antiparticle.

An example of charge symmetry breaking (CSB) occurs in the decay of the $\psi'(3686) \to (J/\psi)\pi^0$. This is forbidden by charge symmetry because $P_{\rm cs}|\pi^0\rangle = -|\pi^0\rangle$, but it occurs because the physical neutral pion system is a mixture of bare $\pi^0$ ($T = 1$), $\eta$, and $\eta'$ (both $T = 0$) mesons that are mixed by small electromagnetic effects and much larger effects of the light quark mass difference.

If charge symmetry were exact, the proton and the neutron would have the same mass. CSB causes the neutron to be about 0.1% heavier than the proton. The electrostatic repulsion between quarks should make the proton heavier. But the mass difference between the quarks wins over their electrostatic repulsion by a factor of about two, making it the dominant cause of CSB.

The neutron-proton mass difference has important consequences for the structure of the universe because it means that a neutron can decay into a proton (plus an electron and an anti-neutrino) in radioactive beta decay. When protons and neutrons combined to form elements in the first few minutes after the Big Bang, the resulting elemental abundances depended on the neutron-proton mass difference and the lifetime of the neutron. All the neutrons that survived were bound inside nuclei, which left many protons free. The interactions between these protons are the main source of energy in stars like the Sun.

The mass difference between the up and down quarks is the *only* strong interaction effect that breaks charge symmetry (1) and isospin invariance. Electromagnetic effects also violate charge symmetry, so that these effects and those of the quark mass difference are responsible for all charge-dependent effects and all violations of isospin. However, to focus on the light quark mass difference it is necessary to concentrate on CSB. For example, the strong interaction scattering lengths in the $T = 1$ neutron-proton ($\sim -24$ fm) and neutron-neutron ($\sim -19$ fm) systems are different. This arises mainly from the mass difference between the charged and neutral pions, which is itself dominated by electromagnetic effects. Thus a seemingly strong interaction scattering length difference really has its origins in electromagnetism. Another example is that the ratio of cross sections, $\sigma(pp \to d\pi^+)/\sigma(pn \to d\pi^0) = 1/2$, is true only if isospin invariance holds, giving the $pn$ system a 50% probability of having an isospin of unity. Electromagnetic effects such as the initial and final state Coulomb interactions and the $\pi^+ - \pi^0$ mass difference (which overwhelms the influence of the CSB nucleon mass difference) cause deviations from the value of 1/2. The examples presented here allow us to state that all charge dependence (or isospin violation) that is *not* CSB is dominated by electromagnetic effects. If one wants to study the consequences of the up-down quark mass difference it is necessary to consider only CSB effects.

A review article is timely now because of recent experimental progress (2; 3) in measuring CSB in the production of neutral pions in neutron-proton and deuteron-deuteron collisions. The current importance of such studies is increased by new theoretical insight – an appropriate use of a convergent effective field theory (EFT) can be equivalent to using QCD. Indeed, the last decade has seen a paradigm shift in trying to understand CSB from within a meson-exchange framework to within an effective field theory framework. In the EFT scheme, the degrees of freedom are nucleons and pions, and precise calculations of strong interaction effects are possible, provided expansions in terms of small momenta converge. Chiral symmetry is used to constrain the forms of terms that incorporate the underlying light quark mass difference and electromagnetic effects that cause all CSB effects. The use of different experiments combined with a carefully constructed theory of the electromagnetic effects should allow the isolation of the quark mass difference effect.

## 1.1 Modern Manifestations

Three diverse topics of high current interest are related to CSB.

### 1.1.1 EXTRACTION OF STRANGENESS FORM FACTORS

The discovery that valence quarks carry only a small fraction of the nucleon spin (4), and the resulting intensified search for strangeness in the nucleon, has brought attention to understanding the role of nucleonic CSB. In principle any form factor of the nucleon receives three independent contributions from $u, d$ and $s$ quarks (neglecting the far smaller contributions of $b$, $c$, and $t$ quarks). If charge symmetry holds, the $u, d$ contributions to neutron form factors are equal to the $d, u$ contributions to proton form factors. The number of independent contributions is thereby reduced to two and thus measurements of the parity violating left-right asymmetry in electron-proton scattering can determine form factors whose origin can lie only in the strange and anti-strange quarks of the nucleon (5; 6; 7; 8) The corrections caused by CSB can be estimated using quark models and are small (9), but may be detectable in future measurements.

### 1.1.2 THE NUTEV ANOMALY

The NuTeV group (10) measured charged and neutral current weak reactions for deep inelastic scattering of neutrinos and anti-neutrinos by iron targets. Ratios of cross sections can be used to determine the weak mixing angle, provided the target is isocalar, the nuclear strangeness content can be ignored, charge symmetry holds, and a variety of nuclear effects can be neglected. With these provisos, the Paschos-Wolfenstein (PW) (11) relation holds for the Standard Model: $\left(\langle\sigma_{NC}\rangle^\nu - \langle\sigma_{NC}\rangle^{\bar\nu}\right)/\left(\langle\sigma_{CC}\rangle^\nu - \langle\sigma_{CC}\rangle^{\bar\nu}\right) = \frac{1}{2} - \sin^2\theta_W$, where $\langle\sigma_{NC}\rangle^{\nu,\bar\nu}$ and $\langle\sigma_{CC}\rangle^{\nu,\bar\nu}$ are the neutrino and anti-neutrino neutral-current and charged-current inclusive total cross sections, averaged over the proton and neutron. This simple relation follows from equating the $u, d$ contributions from protons with those of the $d, u$ contributions from neutrons, and then taking the difference between the $\nu$ and $\bar\nu$ cross sections to isolate the neutral-current axial-vector interference term proportional to $\frac{1}{2} - \sin^2\theta_W$. The NuTeV value for $\sin^2\theta_W$ is three standard deviations larger than the Standard Model prediction, which is normalized to other precise internally-consistent electroweak measurements near the $Z^0$ pole. However CSB causes a change in the PW relation (12; 13). This essentially model-independent correction removes about half of the discrepancy in $\sin^2\theta_W$ (14).

### 1.1.3 HADRONIC CORRECTIONS TO $g-2$

The anomalous magnetic moment of the muon $a_\mu \equiv (g-2)/2$ is experimentally known to an amazing relative precision of 0.5 parts in a million (15). The measured result is either 0.9 or 2.4 standard deviations higher than predicted by standard model theory (16). The two limits are derived from using different methods to obtain the hadronic vacuum polarization term (17). Although the lowest-order hadronic vacuum polarization contribution can be evaluated in terms of the experimentally measured cross section for $e^+e^- \to$ hadrons using a dispersion integral representation (18), $\tau$ decay of very high accuracy can also be used to obtain hadronic input that is directly related to the isovector photon vacuum polarization currents if isospin invariance and unitarity hold. The corrections arising from the light quark mass difference and unknown isospin violating decays are significant at the required level of accuracy; but even with these corrections, the $e^+e^-$ and $\tau$ data are incompatible (17). The hadronic $\tau$ decay data leads to the smaller difference with theory. A key point in relating the $\tau$ and $e^+e^-$ data involves accounting accurately for the CSB effects of $\rho^0 - \omega$ mixing (19). This correction to the anomalous magnetic moment is about 20% of the difference between the methods.

## 2  Phenomenology

The basic mechanisms and related critical experimental facts are discussed here.

### 2.1  Mechanisms of Charge Symmetry Breaking

In the Standard Model the breaking of charge independence and charge symmetry occurs via the non-zero value of $m_d - m_u$ and via quark electromagnetic effects. But quarks are confined, so the task of relating the underlying theory to experimental observations necessarily involves theoretical procedures. Using non-perturbative QCD to compute the small consequences of the quark mass differences accessible in low to medium energy experiments is not yet possible without employing models. The venerable meson-exchange theory of nuclear forces can be used to include charge dependent effects successfully because such effects can often be added without introducing new parameters to describe the strong interaction. More recently, effective field theory (EFT) has been used to relate QCD to the nucleon-nucleon (NN) interaction, to justify the use of meson exchange terms, and to obtain effective operators that violate charge independence. Using EFT can be equivalent to using QCD, so applying this new theoretical tool to understand CSB is a major theme of the present review.

2.1.1  MASS SPLITTINGS OF HADRONIC ISOSPIN MULTIPLETS   The incorporation of quarks into our thinking has led to the emergence of a common explanation of the mass differences within isotopic multiplets that has resolved (1) an old and difficult issue: "the sign of the isotopic mass splitting puzzle". Before quark physics, the only definite mechanism for the breaking of charge independence was the electromagnetic interaction, but naive estimates and sophisticated evaluations based on the electromagnetic interaction gave the inevitable result that a charged particle (proton) should be heavier than its neutral partner (neutron). But actual masses differ from this scheme: $m_n > m_p$ and $m_{K^0} > m_{K^-}$, while $m_{\pi^0} < m_{\pi^+}$; furthermore $m_{\Sigma^+} < m_{\Sigma^0} < m_{\Sigma^-}$. This riddle is solved by realizing that we should not arrange multiplets by their electric charge, but by their $u - d$ flavor. If two hadrons are related by replacing a $u$ quark by a $d$ quark, the $d$-rich system is heavier. This holds for all known mesons and baryons (20), except perhaps for the $\Sigma_c$ masses that depend on enhanced electromagnetic effects.

2.1.2  MIXING HADRONS   If isospin invariance holds, the neutral mesons of $u - d$ flavor are states of pure isospin, given schematically as

$$|I=1\rangle = \frac{1}{\sqrt{2}}|u\bar{u}\rangle - \frac{1}{\sqrt{2}}|d\bar{d}\rangle, \qquad |I=0\rangle = \frac{1}{\sqrt{2}}|u\bar{u}\rangle + \frac{1}{\sqrt{2}}|d\bar{d}\rangle. \qquad (4)$$

The isospin of a state is determined by the final states obtained via strong decay processes, i.e., $2\pi$ for $I=1$ and $3\pi$ for $I=0$. However, the perturbing effects of the quark mass difference and electromagnetic effects cause the states to mix. For example, the quark mass contribution to the QCD Hamiltonian, $H_m = m_d \bar{d}d + m_u \bar{u}u$, gives a mixing matrix element of the form

$$\langle I=1|H_m|I=0\rangle = m_u - m_d, \tag{5}$$

which is negative. Electromagnetic effects also enter, as we shall discuss.

Neutral mesons are mixtures of $I=0$ and $I=1$ states, with the largest mixing occurring for two nearly degenerate states. The best studied dramatic example is that of $\rho^0 - \omega$ mixing, now known as an important element in understanding the hadronic vacuum polarization contribution to the muon anomalous magnetic moment. The effects of this matrix element were observed long ago (21) in the annihilation process $e^+e^- \to \pi^+\pi^-$. The interference between the effects of $2\pi$ emission from the physical $\rho$ and $\omega$ is manifest as a huge shoulder on the $\rho$ peak of the cross section. The magnitude, phase, and dependence on $s$ (square of the four-momentum of the virtual meson) of the $\rho^0 - \omega$ complex self-energy matrix $\Pi_{\rho\omega} \approx \langle \rho^0|H|\omega\rangle$ have been extracted (22) using a rigorous theoretical framework to analyze $e^+e^-$ cross sections (instead of the branching ratio (23)). The off-diagonal matrix elements are

$$\Pi_{\rho\omega}(m_\omega^2) = -3500 \pm 300 \ - (300 \pm 300)i \text{ MeV}^2, \Pi'_{\rho\omega}(m_\omega^2) = 0.03 \pm 0.04, \tag{6}$$

where $\Pi'_{\rho\omega}$ characterizes the $s$-dependence of $\Pi_{\rho\omega}(s)$ about $s = m_\omega^2$. Both the imaginary part and $s$-dependence of $\Pi_{\rho\omega}(s)$ are statistically consistent with zero. The specific results of Eq. (6) are derived from early data, but newer data exist (24). The most recent analysis using fits obtained from (19) yields

$$\Pi_{\rho\omega}(m_\omega^2) = -4290 \pm 360 \text{ MeV}^2 \tag{7}$$

with a phase varying between $13.4°$ and $18.1°$. The spread in values includes the effects of model dependence.

The matrix element $\langle \rho^0|H|\omega\rangle$ includes the effect of the electromagnetic process $\omega \to \gamma \to \rho$. The $\omega$ and $\rho$ electromagnetic decay matrix elements are well known, and Ref. (25) gives $\langle \rho^0|H_{em}|\omega\rangle = 640 \pm 140$ MeV$^2$. This electromagnetic matrix element influences NN scattering as part of the one photon exchange term, and does not enter in the short-ranged CSB strong interaction. Only the strong contribution ($H = H_S + H_{em}$) given by (if Eq. (7) is used)

$$\langle \rho^0|H_S|\omega\rangle = -4930 \pm 390 \text{ MeV}^2 \tag{8}$$

enters. The units are mass-squared because the self-energy appears in the mesonic Klein-Gordon equation. Removing the covariant normalization allows the mixing matrix element to be expressed as $\langle \rho^0\|H\|\omega\rangle = \langle \rho^0|H|\omega\rangle/(m_\rho + m_\omega) \approx -3.1 \pm 0.25$ MeV, comparable with expectations based on Eq. (5).

The $\rho^0 - \omega$ mixing matrix element has been the subject of much discussion because of its role in the CSB NN interaction and the related importance in nuclear physics. Two questions have been raised: (1) is the historically common neglect of direct CSB emission of two pions from the $\omega$ accurate (26)? (2) how does this matrix element depend on $s$ (27)?

The term direct CSB emission refers to a bare isospin-0 $\omega$ meson emitting two pions via a CSB operator. For example, the Weinberg-Tomazowa interaction connects a vector meson to the vacuum with a term that depends on $m_d - m_u$. The small, but non-vanishing phase of the mixing matrix element (26) indicates that there is non-vanishing direct emission. The implications of this for CSB in NN interactions have not been worked out.

We now turn to the $s$ dependence. For electron-positron annihilation $s \approx m_\rho^2$, but for NN scattering $s < 0$. Various authors (see the review (28)) argued that there is significant variation such that $\langle \rho^0 | H_S | \omega \rangle$ vanishes at $s = 0$, and therefore the resulting CSB NN potential is obliterated. However the natural consequences of such a variation (29; 30) are that a strongly $s$-dependent $\rho^0 - \omega$ mixing implies a strong variation of the $\rho - \gamma^*$ transition matrix in contrast with observation (31). Later work (32), using a model in which vector mesons are described as nucleon–anti-nucleon pairs, obtained a vector dominance model that reduced the $s$-dependence of $\rho - \gamma^*$ transition matrix. This model is fine tuned. Replacing the nucleon mass by a slightly smaller value leads to a large variation with $s$.

Fundamental reasoning expects little variation with $s$. New EFT treatments of the CSB NN interaction estimate a term of the size obtained from on-shell mixing (33) in accord with the venerable tadpole dominance picture (34). See also Ref. (35). But perhaps the most important reason for small variation is the data, as summarized by the zero value of $\Pi'_{\rho\omega}$ in Eq. (6). There is no viable model that accommodates $\langle \rho^0 | H_S | \omega \rangle = 0$ at $s = 0$.

Meson mixing causes the charge dependence of the NN potential when a $T = 0$ meson ($\omega$ or $\eta$) emitted from one nucleon fluctuates into a $T = 1$ meson ($\rho^0$ or $\pi^0$) that is absorbed on another nucleon. The resulting interaction also depends on the product of the two strong coupling constants that controls emission and absorption. We shall see that the $\eta$ nucleon coupling constant, $g_{\eta NN}$, plays a vital role in the CSB production of neutral pions, so we review developments concerning its magnitude. An early analysis (36) using one-boson exchange potentials in NN scattering gave $g_{\eta NN}^2/4\pi = 3.86$, but the data show little sensitivity to $\eta$ exchange and high-accuracy fits can be achieved (37) using $g_{\eta NN}^2/4\pi = 0$. Indeed, the possibility of a vanishing coupling constant was raised earlier. The detailed analysis of NN total cross sections and $p\bar{p}$ data using dispersion relations (38) found that $g_{\eta NN}^2/4\pi = 0$. This is consistent with extractions from the nucleon pole in the amplitude $\pi N \to \eta N$ yield (39) that give $0.5 > g_{\eta NN}^2/4\pi \geq 0$. Photoproduction reactions on a nucleon (40) yield (see their Fig. 2) the very small value $g_{\eta NN}^2/4\pi = 0.1$. Here we use the range

$$0.10 \leq \frac{g_{\eta NN}^2}{4\pi} \leq 0.51, \tag{9}$$

where 0.51 was used in (41) as a small coupling constant, but now must be regarded as a large coupling constant. Ref. (41) used $SU(3) \times SU(3)$ chiral symmetry to argue that the phases of the $\pi-$ and $\eta-$nucleon coupling constants are the same, but this needs to be re-examined because of the failure to understand the magnitude of $g_{\eta NN}$.

2.1.3 CHARGE DEPENDENT MESON-NUCLEON COUPLING CONSTANTS  In principle, charge dependent coupling constants where $g_c \neq g_0$ ($c$ refers to a charged meson) arise from photon loop diagrams, from vector-meson mixing in loop diagrams, and from the nucleon mass difference. Note that QED effects do not contribute to $g_c - g_0$ if an appropriate renormalization procedure is used (42).

2.1.4 CHARGE SYMMETRY BREAKING QUARK CONDENSATES AND QCD SUM RULES In a chiral version of QCD, with $N_f$ vanishing quark masses there is an exact chiral $\mathrm{SU}(N_f) \times \mathrm{SU}(N_f)$ symmetry. The QCD vacuum breaks this symmetry to $SU(N_f)$ by forming non-vanishing chiral condensates. Corrections for the up and down quark masses can be incorporated using chiral perturbation theory, and the related condensates differ. This difference can be used with the QCD sum rule technique to account for the charge dependence of the $u, d$ constituent quark masses and thereby reproduce the neutron-proton mass difference and CSB mesonic decays (43).

## 2.2 Classification of Charge Dependent Nucleon-Nucleon Forces

We have seen a host of charge-dependent and charge asymmetric mechanisms. Understanding the sensitivities of different nucleon-nucleon experiments to different fundamental mechanisms is aided by characterizing the charge dependence of nuclear forces according to their isospin dependence. The discussion of Henley and Miller (44; 45) listed four classes of forces between two nucleons, $i$ and $j$.

Class (I): Forces that are isospin or charge independent. Such forces, $V_I$, obey $[V_I, \mathbf{T}] = 0$ and thus have the isoscalar form

$$V_I = a + b\boldsymbol{\tau}(i) \cdot \boldsymbol{\tau}(j) \tag{10}$$

where $a$ and $b$ are Hermitian isospin-independent operators.

Class (II): Forces that maintain charge symmetry but break charge independence. These can be written in an isotensor form

$$V_{II} = c[\tau_3(i)\tau_3(j) - \frac{1}{3}\boldsymbol{\tau}(i) \cdot \boldsymbol{\tau}(j)]. \tag{11}$$

The Coulomb interaction leads to a Class II force as do the effects of the pion mass difference in one pion and two pion exchange interactions and the possible effects of charge-dependent coupling constants.

Class (III): Forces that break both charge independence and charge symmetry, but which are symmetric under the interchange $i \leftrightarrow j$ in isospin space,

$$V_{III} = d[\tau_3(i) + \tau_3(j)], \tag{12}$$

where $d$ is a Hermitian operator that is symmetric under the interchange $\boldsymbol{\sigma}(i) \leftrightarrow \boldsymbol{\sigma}(j)$. A class III force differentiates between $nn$ and $pp$ systems. However, it does not cause isospin mixing in the two-body system, because $[V_{III}, T^2] \propto [T_3, T^2] = 0$. This force vanishes in the $np$ system. The effects of the exchange of a mixed $\rho^0 - \omega$ meson yield a significant class III force as do the effects of the neutron-proton mass difference in the two pion exchange potential.

Class (IV): Forces that break charge symmetry and therefore charge dependence. These forces cause isospin mixing. They are of the form

$$V_{IV} = e[\boldsymbol{\sigma}(i) - \boldsymbol{\sigma}(j)] \cdot \mathbf{L} \, [\tau_3(i) - \tau_3(j)] + f[\boldsymbol{\sigma}(i) \times \boldsymbol{\sigma}(j)] \cdot \mathbf{L} \, [\vec{\tau}(i) \times \vec{\tau}(j)]_3, \tag{13}$$

where $e$ and $f$ are spin-independent operators. The $e$ term receives contributions from $\gamma$ and $\rho - \omega$ exchanges while $f$ is caused by the influence of the nucleon mass difference on $\pi$ and $\rho$ exchange. Class IV forces vanish for $nn$ and $pp$, but cause spin-dependent isospin mixing effects in the $np$ system.

## 2.3 Nucleon-Nucleon Scattering

We discuss charge independence breaking of the $^1S_0$ NN scattering lengths to illustrate the consequences of the above mentioned mechanisms and to review the data.

Charge independence, $[H_S, \mathbf{T}] = 0$, imposes the equalities of the nucleon-nucleon scattering lengths $a_{pp} = a_{nn} = a_{np}$. But electromagnetic effects are large and it is necessary to make corrections. These corrections are discussed in Refs. (1) and (30). There have been no major recent changes, but new work on the $nn$ scattering length is reviewed in Sect. 2.3.1. The results are

$$\begin{aligned} a^N_{pp} &= -17.3 \pm 0.4 \text{ fm} & r^N_{pp} &= 2.85 \pm 0.04 \text{ fm} \\ a^N_{nn} &= -18.8 \pm 0.3 \text{ fm} & r^N_{nn} &= 2.75 \pm 0.11 \text{ fm} \\ a^N_{np} &= -23.77 \pm 0.09 \text{ fm} & r^N_{np} &= 2.75 \pm 0.05 \text{ fm} \end{aligned} \quad (14)$$

in which the superscript $N$ represents the "nuclear" effect obtained after the electromagnetic corrections have been made. Experimental errors for the $pp$ and $np$ systems are negligible, the error bar is dominated by the model dependence introduced by removing the electromagnetic effects. The differences among these scattering lengths represent charge independence and charge symmetry breaking.

The large percentage differences among the scattering lengths may seem surprising, but it is the inverse of the scattering length that is related to the potential. For nucleons of mass $M$ and two different potentials, $\Delta V = V_1 - V_2$, manipulating the Schroedinger equation tells us that the scattering lengths, $a_1$ and $a_2$, are related by

$$\frac{\Delta a}{a} = -aM \int_0^\infty dr\, u^2(r) \Delta V(r), \quad (15)$$

where $\Delta a = a_2 - a_1$, $a^2 = a_1 a_2$, and $u$ is the corresponding zero energy wave function, normalized to approach $1 - r/a$ as $r$ approaches $\infty$ and $u(0) = 0$. The effects of changing the potential on the wave function are very small, so that this formula is accurate (46) to about 2% in $\Delta a$. The large s-wave scattering lengths strongly enhance the influence of $\Delta V$, and one can show (47) that

$$\frac{\Delta a}{a} = -(10 - 15)\frac{\Delta V}{V}, \quad (16)$$

where a simple shape is chosen for the potentials $V_{1,2}$. The variation between 10 and 15 arises from using different radial shapes for V(r). In contrast the changes in the effective range are negligible. One defines $\Delta a_{CD}$ to measure the charge independence breaking, with

$$\Delta a_{CD} \equiv \frac{1}{2}(a^N_{pp} + a^N_{nn}) - a^N_{np} = 5.7 \pm 0.3 \text{ fm}. \quad (17)$$

This charge independence breaking is caused by a Class II force that may be estimated [using Eq. (16)] to be about $\Delta V/V \sim 2.5\%$. The breaking of charge symmetry, caused by Class III forces, is represented by

$$\Delta a_{CSB} \equiv a_{pp}^N - a_{nn}^N = 1.5 \pm 0.5 \text{ fm}, \tag{18}$$

and amounts to about a 0.6% effect.

The values of $\Delta a_{CD}$ and $\Delta a_{CSB}$ above have been well understood in terms of meson exchange models (1; 30). The charge dependence is dominated by the effects of the pion mass difference in the one and two pion exchange potentials. It is necessary to include an estimate (46) of the effects of $\pi\gamma$ exchange to obtain a quantitative understanding of $\Delta a_{CD}$. The value of $\Delta a_{CSB}$ was explained entirely in terms of $\rho^0 - \omega$ mixing (23), provided a sufficiently large $\omega$NN coupling constant is used. More recent work is discussed in Sect. 3.3.

2.3.1 NEUTRON-NEUTRON SCATTERING LENGTH   The lack of a free neutron target prevents a straightforward observation of neutron-neutron scattering. Therefore the most reliable determinations of $a_{nn}$ and $r_{nn}$ occur in three-body reaction studies with only two strongly-interacting particles in the final state. Consequently $a_{nn}$ and $r_{nn}$ are mainly deduced in studies of the reaction $\pi^- d \to nn\gamma$.

The first precise measurements of the neutron-neutron scattering lengths in the $\pi^- d \to nn\gamma$ reaction made at PSI (48; 49) indicated that the $^1S_0$ $nn$ force was significantly more attractive than the $pp$ force, and this led to Eq. (14). A more recent experiment (50) found $a_{nn} = -18.50 \pm 0.05(\text{stat.}) \pm 0.44(\text{syst.}) \pm 0.30(\text{theory})$ fm. (There is a ($-0.3$ fm) correction to $a_{nn}$ caused by the electromagnetic interaction (1), so this result is consistent with that of Eq. (14).) The extraction was obtained by fitting the shape of the neutron time-of-flight spectrum using the GGS reaction model (51). The theoretical error of 0.3 fm was dominated by uncertainties in the pion scattering wave function. The theoretical work for the PSI results compared the GGS model with work done by de Téramond and collaborators (48). The latter used a dispersion relation approach for the final state interaction with a theoretical error of the order 0.3 fm, similar to GGS. The techniques of effective field theory have been used recently by Gardestig and Phillips to reassess the theoretical uncertainty (52). Chiral power counting gives a clearly defined procedure to estimate the theoretical error and provides a systematic and consistent way to improve the calculation if needed. The result is a theoretical uncertainty of $\pm 0.2$ fm, provided $a_{nn}$ is extracted using data taken at kinematics where final state interactions are dominant. This confirms the conclusion of GGS from thirty years ago.

A recent $nd$ break-up experiment reports $a_{nn} = -16.1 \pm 0.4$ fm (53), more than five standard deviations from the standard value of Eq. (14). The result of Ref. (53) is also in disagreement with another $nd$ experiment that claims $-18.7 \pm 0.6$ fm (54). Our view is that the $nd$ break-up is fraught with complications (30) in deciphering the influence of various two- and three-body final state interactions. Additionally, using $a_{nn} = -16.1$ fm instead of $-18.5$ would destroy the currently accepted explanation of the Nolen-Schiffer anomaly (Sect. 2.5), so it is important that future three-body experiments confront this difference.

## 2.4   Observation of Class IV Forces

This subsection describes the successful search for the Class IV forces of Eq. (13) that cause spin-dependent isospin mixing effects in the $np$ system. As a result the analyzing power of polarized neutrons scattered from unpolarized protons, $A_n(\theta_n)$, differs from the analyzing power of polarized protons scattered from unpolarized neutrons, $A_p(\theta_p)$ (45; 55). Measurements at TRIUMF [477 MeV (56) and 347 MeV (57)] compared analyzing powers with a neutron beam and a proton target alternately polarized. A similar experiment at IUCF observed $A_n$ and $A_p$ with both the neutron and the proton initially polarized at 183 MeV (58). Determining the CSB observable $\Delta A = A_n - A_p$ from measurements of $A_n$ and $A_p$ when both are large would have required measurement of the average beam or target polarization to a few hundredths of a percent, which is not possible to this day. However, these analyzing powers pass through zero at one angle, $\theta_0$, for each of the energies of the TRIUMF and IUCF experiments. If $\theta_0$ for polarized neutrons differs from $\theta_0$ obtained for polarized protons, then $\Delta \theta \equiv \theta_0(n) - \theta_0(p) \neq 0$ and charge symmetry has been violated. The results of the three beautiful experiments presented in terms of $\Delta A$ [$= (dA/d\theta) \Delta \theta$], are summarized in Ref. (57). $\Delta A/A$ varies from about $30 \times 10^{-4}$ at 183 MeV to about $60 \times 10^{-4}$ at 347 and 477 MeV. These results are very well understood in terms of CSB meson exchange mechanisms (59; 60; 61): small $\gamma$ effects, $n - p$ mass difference in $\pi - \rho$ exchange, and $\rho^0 - \omega$ mixing. A pion exchange effect arising from the presence of the $n - p$ mass difference in the evaluation of the vertex function dominates the 477 MeV measurement (59). The $\rho^0 - \omega$ mixing term has a significant but non-dominating influence at 183 MeV. The observation of non-vanishing shifts in $\theta_0$ represents the first discovery of CSB in a system that has no Coulomb effects. The TRIUMF and IUCF $n - p$ scattering experiments increased the awareness of the fundamental mechanism for CSB in the NN interaction. This led to a resolution of the long-standing Nolen-Schiffer anomaly.

## 2.5 Mirror-Nuclei Binding Energies

Charge symmetry breaking is manifest in the differences between mirror nuclei $(N, Z) = (Z', Z'+1)$ and $(Z'+1, Z')$. This is the natural generalization of the $n-p$ mass difference. Finding the influence of the light-quark mass difference involves removing electromagnetic effects and their interference with the strong interaction in a many-body system. Furthermore the connection with the underlying NN interaction could not occur until the value of $a_{nn}$ was established.

2.5.1 THE $^3$HE-$^3$H BINDING ENERGY DIFFERENCE  The ground state binding energy difference $\Delta B = B(^3\text{H}) - B(^3\text{He}) = 764$ keV is a measure of CSB (62). The proton rich $^3$He nucleus is less deeply bound because of the repulsive influence of the Coulomb interaction and other electromagnetic effects. Such effects must be removed to determine the strong interaction CSB. The three body system is the best for such evaluations because the most important electromagnetic terms can be evaluated in a model independent way using measured electromagnetic form factors (63). Furthermore, exact three body calculations are routine. In 1987 Coon and Barrett (23) used the then recent electron elastic form factor measurements to extract $\Delta B = 693 \pm 19 \pm 5$ keV, where the first uncertainty is due to the determination of the form factors, and the second to the small model dependence of some relativistic effects. Similar values of $\Delta B$ were obtained by solving the Faddeev equation in Ref. (64). The CSB difference is about 71 keV, to be accounted for by the CSB of the strong interaction. The use of a $\rho^0 - \omega$ exchange potential which reproduces $\Delta a_{CSB}$ yields good agreement with the experimental difference. However, any other potential that can account for $\Delta a_{CSB}$ such as that arising from the nucleon and $\Delta$ mass difference (65; 66) will also account for the binding energy difference.

### 2.5.2 NOLEN–SCHIFFER ANOMALY AND OTHER NUCLEAR STRUCTURE EFFECTS

The CSB seen for A=3 also occurs for mirror nuclei. Nolen and Schiffer (67) made a detailed analysis of the binding energy differences of mirror nuclei, finding electromagnetic effects too weak to understand the data. Instead there is a residual effect due to the CSB of the strong interaction: the neutron rich nuclei were seen to be more deeply bound (by about 7%) than the proton rich nuclei. Including additional detailed nuclear structure effects reduced that number, but only to a rather substantial 5% (1; 68; 69). Negele suggested (68) that the CSB in the NN force could be responsible for this 5%.

Efforts to better understand the anomalous 5% were initially thwarted by a lack of knowledge of the neutron-neutron scattering length. This has been settled for some time and several authors starting with Ref. (70) showed that using a nucleon-nucleon interaction that accounts for $\Delta a_{CSB}$ (and dominated by the effects of $\rho^0 - \omega$ mixing) accounts for the bulk of the Nolen-Schiffer anomaly.

The most recent work is that of Ref. (65) which investigates three models for the CSB of the NN interaction (based upon $\rho^0 - \omega$ mixing, nucleon mass splitting, and phenomenology) that reproduce the empirical values of the $^1S_0$ scattering lengths. Each model yields very different CSB for the $^3P_J$ phase shifts, and the consequent differences in mirror nuclei are studied for $A \geq 3$. The $^3$H-$^3$He binding energy difference is dominated by $\Delta a_{CSB}$, but binding energy differences of heavier nuclei receive about 50% of their contribution from NN partial waves beyond $^1S_0$. Consequently, the predictions by the various CSB models differ by 10 to 20%. Even so, using any of the CSB potentials reduces the anomaly to a small fraction (less than one percent) of the binding energy difference.

The common feature (30) of all models is that the CSB of the strong interaction, as driven by the light quark mass difference, removes the anomaly. The Nolen-Schiffer anomaly is no longer a puzzle.

Other old puzzles concerning the energy differences between heavier nuclei ($^{48}$Ca, $^{90}$Zr, $^{208}$Pb) and their isobaric analog states are reviewed in (30). The use of both charge symmetry and charge independence breaking forces consistent with NN data is essential in understanding the remaining binding energy differences. A new example, Ref. (71), found that isovector and isotensor energy differences between yrast states of isobaric multiplets in the lower half of the $pf$-shell region are quantitatively reproduced using the shell model. The isospin non-conserving nuclear interactions are at least as important as the Coulomb potential.

### 2.5.3 CHARGE SYMMETRY BREAKING IN HYPERNUCLEI

Hypernuclear binding energy differences allow tests of charge symmetry of the $\Lambda$N interaction. The main source of information is the mirror pair, $^4_\Lambda$H and $^4_\Lambda$He (72; 73). The binding energy differences of mirror pairs are not understood. Recent progress has involved the ability to make exact calculations for few-baryon systems, but Nogga *et al.* (74) find that none of the existing interactions correctly binds all four-body hypernuclei states and plan to constrain hyperon-nucleon interactions using the four-body separation energies.

## 3  Effective Field Theory

Effective field theory is the modern tool for handling strong interactions among hadrons. One may obtain the results of QCD by using a chiral Lagrangian ex-pressed in terms of hadronic degrees of freedom (75; 76; 77; 78). This surprising possibility derives from an argument that the most general Lagrangian that respects unitarity, has correct properties under cluster decomposition, and has the same symmetries as QCD should be equivalent to QCD. Predictive power is obtained if an expansion in momentum, formulated using power counting arguments, can be shown to converge. We discuss those aspects of EFT that are most relevant for understanding CSB, relying heavily on the reviews (77; 78).

For nuclear physics the principal aim and gain of using EFT is to handle the short distance physics. In the traditional approach one resorts to modeling short-distance interactions ($\lesssim 0.7$ fm) between two or more nucleons. EFT offers a systematic procedure, that of treating ultraviolet-divergent integrals that appear in evaluating loop diagrams by using cutoffs at a scale $\Lambda$ and counter terms with magnitudes expressed in terms of low energy constants (LECs) that depend implicitly on $\Lambda$. If the Lagrangian and renormalization procedure is correctly constructed, the computed observables are independent of $\Lambda$. Then one computes low-energy observables, measured at a small momentum scale $Q$ ($Q \ll \Lambda$) as an expansion in powers of $Q$. If only a finite number of LECs contribute at a given order in the expansion, one may obtain a Lagrangian with a small number of coefficients to be determined from experimental data, fundamental theory, or quark models. The Lagrangian can then be used to predict other observables. If pions are included explicitly, observables are expressed as a series in $Q/M_{\text{QCD}}$ where $M_{\text{QCD}} \sim 4\pi f_\pi$ (the pion decay constant $f_\pi \simeq 92.6$ MeV) is the baryon mass scale. A computation is regarded as converged if increasing the order in $Q/M_{\text{QCD}}$ of a calculation increases its precision without introducing too many new LECs. The term "power counting" is used to denote the technique of connecting the order of the expansion in powers of $Q/M_{\text{QCD}}$ and the necessary LECs.

## 3.1 Basics

To understand CSB in hadronic physics we use a Lagrangian, $\mathcal{L}$, expressed in terms of pions, nucleons, $\Delta$'s, and photons. Heavier mesons are also included to account phenomenologically for multi-pion interactions. The specific connection between QCD and $\mathcal{L}$ is approximate chiral symmetry, which determines how pions interact. QCD has a chiral $SU(2)_L \times SU(2)_R$ symmetry, if one takes the chiral limit of neglecting the masses $m_u$ and $m_d$. Chiral symmetry is assumed to be broken spontaneously to its diagonal subgroup, the $SU(2)_{L+R}$ of isospin. Massless Goldstone bosons (79), naturally identified as pions, are associated with the three broken generators, and their fields $\boldsymbol{\pi}$ exist on a three-sphere $S^3$, of radius $f_\pi$ that represents the set of possible vacua. The interactions of pions are invariant under chiral rotations, and therefore must involve derivatives on $S^3$ combined with specific non-linear terms.

As long as the quark masses are small enough, their only effect is to change this picture slightly. The quark mass difference breaks charge symmetry. In low-energy EFT, the effect of quark-mass terms can be reproduced by including all terms that break chiral symmetry in the same way. Such interactions contain powers of $m_d \pm m_u$ times factors involving $\boldsymbol{\pi}$ without derivatives.

The chiral Lagrangian $\mathcal{L}$ can be constructed using non-linear representations, and the most general expression contains an infinite number of terms that can be grouped according to the chiral index $\Delta$ (80; 81):

$$\mathcal{L} = \sum_{\Delta=0}^{\infty} \mathcal{L}^{(\Delta)}, \quad \Delta \equiv d + f/2 - 2, \tag{19}$$

where $f$ is the number of fermion fields and $d$ is the sum of the number of derivatives, powers of $m_\pi$ (or $(m_d \pm m_u)^{1/2}$), and powers of the $\Delta - N$ mass difference $\delta m$. Chiral symmetry of the strong interaction provides the constraint that $\Delta \geq 0$.

In computing a specific scattering amplitude one achieves predictive power by ordering contributions in powers of the external three-momenta $Q \sim m_\pi$. In any Feynman diagram each space derivative appearing in an interaction produces a three-momentum in a vertex and therefore counts as $Q$. A complication stems from the presence of heavy particles such as the nucleon together with light particles such as the pion. In any loop, integration over the zeroth component of the four-momentum, $k^0$, involves poles at $\sim Q$ corresponding to external three-momenta and to the mass of the pion; and *shallow* poles at $\sim Q^2/2m_N$ corresponding to external nucleon energies. If a process involves at most one heavy particle ($A = 0, 1$), the contour of integration can be closed to avoid the shallow poles so that $k^0 \sim Q$. Then each time derivative counts as $Q$, each four-momentum integration brings a factor $Q^4$, each pion propagator varies as $Q^{-2}$, and the mass term of a nucleon (or $\Delta$) propagator counts as $1/Q$. The kinetic energy terms are of relative $\mathcal{O}(Q/m_N)$ and are treated as corrections.

This power counting allows the contribution of any diagram amplitude to be expressed (77) as

$$T \propto Q^\nu \mathcal{F}(Q/\Lambda), \tag{20}$$

where $\Lambda$ is a renormalization scale, $\mathcal{F}$ is a calculable function of LECs, and $\nu$ is a counting index. For strong interactions(75)

$$\nu = 4 - 2C - A + 2L + \sum_i \Delta_i, \tag{21}$$

where $C = 1$ is the number of connected pieces, $L$ is the number of loops, and the sum runs over all vertices. Since $L \geq 0$ and $\Delta \geq 0$, $\nu \geq 2 - A \equiv \nu_{min}$. The use of Eq. (20) leads to an expansion in $Q/M_{QCD}$ if all of the LECs have "natural" size (given by powers of $M_{QCD} \sim 1$ GeV $\approx 4\pi f_\pi$).

If a process involves more than one stable heavy particle ($A \geq 2$) the shallow poles can not be avoided, and perturbation theory must be re-ordered (82). Shallow poles represent intermediate states that differ from initial states only by nucleon kinetic energies of order $\sim Q^2/m_N$. This $\mathcal{O}(m_N/Q)$ infrared enhancement relative to intermediate states of energy $\sim Q$ invalidates Eq. (21). A more detailed counting is needed.

Weinberg's proposal [(82), see also (83)] was to split amplitudes into reducible and irreducible parts. Irreducible diagrams, in which typical energies resemble those in ordinary Chiral Perturbation Theory, should satisfy the power counting of Eq. (21) with their sum denoted as the potential $V$. Reducible diagrams are obtained from $V$ by solving the Lippmann-Schwinger equation.

Questions have been raised regarding the consistency of this approach, and alternative counting schemes derived (84). These are discussed in Ref. (77). Ref. (85) discussed an expansion that is equivalent to KSW (84) power counting in the $^1S_0$ channel and to Weinberg power counting in the $^3S_1$–$^3D_1$ coupled channels. The most convergent aspect of each power counting scheme is selected to compute phase shifts.

## 3.2 Potentials and fits to NN data

Much work based on Weinberg's power counting has been devoted to developing an EFT potential that defines and evaluates the important terms. Power counting following Eq. (21) for the EFT potential implies that diagrams with an increasing number of loops should be progressively less important. In leading order, $\nu = \nu_{min} = 0$ and the NN potential is simply static one-pion exchange (OPE) augmented by momentum-independent contact terms. The $\nu = 1$ corrections vanish due to parity and time-reversal invariance. The $\nu = 2$ corrections include short-range corrections arising from one-loop pion dressing of the lowest-order contact interactions, and four-nucleon contact interactions with two derivatives or two powers of the pion mass. Two-pion exchange (TPE) diagrams built out of lowest-order $\pi NN$ (and $\pi N\Delta$) interactions also enter at this order. At $\nu = 3$, TPE diagrams that involve the $\pi\pi NN$ seagull vertices from the $\Delta = 1$ Lagrangian appear. At $\nu = 4$ a host of two-loop diagrams and new contact interactions emerge.

The first EFT calculation of all contributions to the NN potential up to $\nu = 3$ was carried out in Refs. (86; 87), and a theoretically-based potential containing all of the spin-isospin structure of phenomenological models was obtained. While phenomenological potentials (88) have similar short- and long-range structure, chiral symmetry is particularly influential in computing TPE. Ref. (89) showed that both isoscalar central and spin-orbit components of the TPE potential are numerically similar to that produced in the $\sigma$ and $\omega$ exchange models. More recently, the tail of the EFT TPE potential, obtained in the limit of a heavy $\Delta$, was substituted for (the tail of) the one-boson exchange in a Nijmegen phase-shift reanalysis of $pp$ data below 350 MeV (90). A drop in $\chi^2$ was observed, and the $\pi\pi NN$ seagull LECs determined by reproducing the NN data are close to values extracted from $\pi N$ scattering. This confirms the validity of chiral TPE.

The description of phase shifts obtained using a potential computed using $\nu = 3$ is inferior to those of "realistic" potentials that use $40 - 50$ parameters to fit data up to 300 MeV with a $\chi^2$ near 1. But the EFT potential has now been extended to $\nu = 4$ (91; 92; 93) and high accuracy fits have been obtained. One can legitimately claim that all but the shortest-ranged components of the nuclear force are understood from a more fundamental viewpoint. Does replacing the mysteries of the shortest ranged interactions by LECs represent true progress?

## 3.3 EFT and CSB

The mass difference between $u$ and $d$ quarks breaks charge symmetry. Analysis of meson masses shows that the ratio $\varepsilon \equiv (m_u - m_d)/(m_u + m_d) \approx 1/3$, suggesting that charge symmetry might be broken by that amount. But experiments show that charge symmetry is typically broken only at the level of a percent or less.

It is spontaneous breaking of chiral symmetry that causes charge symmetry to be good (94). While explicit chiral-symmetry-breaking effects are present already at index $\Delta = 0$ through the pion mass term, operators generated by the quark mass difference appear only at $\Delta = 1$ through a term that contributes to the nu-cleon mass splitting and to certain pion-nucleon interactions. Therefore charge symmetry and its breaking usually competes with isospin-conserving operators of lower order, and its relative size is not $\varepsilon$ but $\varepsilon(Q/M_{QCD})^n$, where $n$ is a positive integer. Thus isospin is a symmetry of the lowest order EFT even though it is not a symmetry of the underlying theory. The only exception to this occurs in the isoscalar $t$ channel in threshold $\pi N$ scattering (95; 94) where there is no contribution from $\mathcal{L}^{(0)}$, so the charge symmetry conserving and breaking amplitudes each start at the same order. The natural arena to observe such effects is CSB $\pi^0$ production experiments (96).

We now discuss the explicit CSB Lagrangian, exploiting the feature that chiral symmetry makes predictions for effects that stem from the small explicit breaking of chiral symmetry generated by the quark masses. Neglecting strangeness, chiral symmetry is essentially an $SO(4)$ internal symmetry. One can show that the quark-mass-difference term in the QCD Lagrangian behaves under $SO(4)$ as the third component of an $SO(4)$ vector (94). Therefore, in the effective hadronic theory, all isospin-violating interactions generated by the quark mass difference break $SO(4)$ as third components of (tensor products of) $SO(4)$ vectors. The operators of interest here involve the nucleon $N$ and pion $\boldsymbol{\pi}$ fields. The leading isospin-violating term at low energies coming from the quark mass difference is the third component of an $SO(4)$ vector (94):

$$\mathcal{L}_{\text{qm}}^{(1)} = \frac{\delta m_N}{2} \left( N^\dagger \tau_3 N - \frac{1}{2Df_\pi^2} N^\dagger \pi_3 \boldsymbol{\pi} \cdot \boldsymbol{\tau} N \right), \tag{22}$$

where $D = 1 + \boldsymbol{\pi}^2/(4f_\pi^2)$. Furthermore, quark interactions generated by ("hard") photon exchange break $SO(4)$ as the 34 component of an $SO(4)$ antisymmetric rank-2 tensor (94). Thus the low-energy effective theory must include isospin-violating interactions that break $SO(4)$ as 34 components of (tensor products of) $SO(4)$ antisymmetric rank-2 tensors, implying that the leading charge symmetry breaking term (Eq. (19) and $d = 0, f = 2$) is:

$$\mathcal{L}_{\text{hp}}^{(-1)} = \frac{\bar{\delta} m_N}{2} \left( N^\dagger \tau_3 N + \frac{1}{2Df_\pi^2} N^\dagger (\pi_3 \boldsymbol{\pi} \cdot \boldsymbol{\tau} - \boldsymbol{\pi}^2 \tau_3) N \right). \tag{23}$$

The next order terms are

$$\mathcal{L}_{\text{hp}}^{(2)} = \frac{\beta_1 + \bar{\beta}_3}{2f_\pi} N^\dagger \boldsymbol{\sigma} \cdot \boldsymbol{\nabla} \pi_3 N, \tag{24}$$

in which $\beta_1$ [$= \mathcal{O}(\varepsilon m_\pi^2/M_{QCD}^2)$] arises from the influence of the quark mass difference on the $\pi$-nucleon coupling constants, and $\bar{\beta}_3$ [$= \mathcal{O}(\alpha/\pi)$] arises from electromagnetic interactions.

Observe that chiral symmetry links the first terms in Eqs. (22,23), which are contributions to the nucleon mass difference, to the second terms, which are isospin-violating pion-nucleon interactions. These pion-nucleon interactions (referred to below as the CSB seagull terms) represent a significant prediction of QCD. It is difficult to isolate the parameters in $\pi N$ scattering, so Ref. (96) suggested that CSB in pion production could be used instead. The only existing constraint on the two terms, $\delta m_N$ and $\bar{\delta} m_N$, is that $\delta m_N + \bar{\delta} m_N = M_n - M_p =$

1.29 MeV. The Cottingham sum rule can be used (97) to give

$$\bar{\delta} m_N = -(0.76 \pm 0.3) \text{ MeV}, \; \delta m_N - \frac{1}{2}\bar{\delta} m_N = 2.4 \pm 0.3 \text{ MeV}, \qquad (25)$$

if several dynamical assumptions are used. Verifying the theory requires that the two terms, Eq. (22) and Eq. (23), be constrained independently, with the reactions $np \to d\pi^0$ and $dd \to \alpha\pi^0$ offering the only possibility.

The other LECs are not well known either. The pion-nucleon CSB parameter $\beta_1 + \bar{\beta}_3$ is constrained by the Nijmegen phase-shift analysis of the NN scattering data (42) to be $\beta_1 + \bar{\beta}_3 = (0 \pm 9) \times 10^{-3}$ (33). Below we estimate the impact of this interaction by modeling the sum $\beta_1 + \bar{\beta}_3$ using $\pi^0 - \eta$ mixing (33). This is consistent with the bound from NN scattering.

Next we discuss how EFT is used to to analyze the charge-dependence of the NN potential in terms of the classes of Sect. 2.2. To compare the various sources of charge dependence, we note that the size of electromagnetic effects in loops is typically $\sim \alpha/\pi$ which, numerically, is $\sim \varepsilon(Q/M_{QCD})^3$.

The leading charge-dependent interactions in Weinberg's power counting have been derived in Ref. (94). The leading $\nu = 0$ potential is class I. The first charge-dependent effects (in addition to the Coulomb interaction) appears at $\nu = \nu_{min}+1$ in the form of a class II potential arising from the pion mass splitting $[\Delta m_\pi^2 = \mathcal{O}(\alpha M_{QCD}^2/\pi)]$ in OPE. A class III force appears at one higher order, $\nu = \nu_{min}+2$. This arises mainly from the quark mass difference through breaking in the $\pi NN$ coupling [via $\pi^0 - \eta$ mixing, $(\beta_1 = \mathcal{O}(\varepsilon m_\pi^2/M_{QCD}^2)]$ in OPE, from heavy meson exchange, treated here as contact interactions $[\gamma_{s,\sigma} = \mathcal{O}(\varepsilon m_\pi^2/M_{QCD}^4)]$, and from the nucleon mass difference $[\Delta m_N = \mathcal{O}(\varepsilon m_\pi^2/M_{QCD})]$. To this order the isospin-violating nuclear potential is a two-nucleon potential of the form

$$V_{ib} = V_{\text{II}} \left[ (\tau_1)_3 (\tau_2)_3 - \boldsymbol{\tau}_1 \cdot \boldsymbol{\tau}_2 \right] + V_{\text{III}} \left[ (\tau_1)_3 + (\tau_2)_3 \right], \qquad (26)$$

where

$$V_{\text{II}} = -\left(\frac{g_A}{4f_\pi}\right)^2 \frac{\vec{q} \cdot \vec{\sigma}_1 \vec{q} \cdot \vec{\sigma}_2}{(\vec{q}^2 + m_{\pi^0}^2)(\vec{q}^2 + m_{\pi^\pm}^2)} (\Delta m_\pi^2 + \Delta m_N^2), \qquad (27)$$

$$V_{\text{III}} = \frac{g_A \beta_1}{4 f_\pi^2} \frac{\vec{q} \cdot \vec{\sigma}_1 \vec{q} \cdot \vec{\sigma}_2}{\vec{q}^2 + m_\pi^2} - (\gamma_s + \gamma_\sigma \vec{\sigma}_1 \cdot \vec{\sigma}_2). \qquad (28)$$

Class IV forces appear only at order $\nu = \nu_{min} + 3$, showing that experimental prowess (see Sect. 2.4) has overcome the small size of such terms.

The pattern of symmetry breaking in QCD naturally suggests a hierarchy of classes in the nuclear potential (94):

$$\frac{\langle V_{\text{M+I}} \rangle}{\langle V_{\text{M}} \rangle} \sim \mathcal{O}\left(\frac{Q}{M_{QCD}}\right), \qquad (29)$$

where $\langle V_{\text{M}} \rangle$ denotes the average contribution of the leading class M potential. This qualitatively explains not only why isospin is a good approximate symmetry at low energies, but also why charge symmetry is an even better symmetry. It gives the observed isospin structure of the Coulomb-corrected scattering lengths of Ref. (1), Eq. (17), and Eq. (18), with $a_{np} \approxeq 4 \times \Delta a_{CD} \approxeq 4^2 \times \Delta a_{CSB}$.

This formalism has been used to do consistent and systematic calculations of charge-dependent effects. The charge-dependent potential arising from one and two pion exchanges up to $\nu = \nu_{min} + 3$ were computed in Refs. (42; 98). In contrast to previous attempts lacking an EFT framework, these results are invariant under both gauge transformation and pion-field redefinition. The class II OPE component arises from diagrams with all possible one-photon dressings of OPE plus the relevant counterterms (42). The chosen renormalization scheme leads to charge-independent $\pi N$ coupling constants. This $\pi\gamma$ potential has been incorporated in a Nijmegen phase shift reanalysis of $np$ data below 350 MeV (42).

We can use the values for the $\pi NN$ coupling constants determined by the Nijmegen analysis to find that their charge dependence ($\beta_1$) is consistent with zero with an uncertainty comparable to our expectation from dimensional analysis and from $\pi^0 - \eta - \eta'$ mixing (33). Similarly, the two contact interactions ($\gamma_{s,\sigma}$) might be viewed as originating in $\rho^0-\omega$ mixing and pseudovector-meson exchange (in particular from close-lying doublets such as $a_1 - f_1$) (33; 99).

The charge dependence of the TPEP arises from two sources. One is the pion mass difference ($\Delta m_\pi^2$) in TPE that generates a class II potential (98); the other is a $\pi\pi NN$ seagull that arises as a chiral partner of the nucleon mass difference ($\Delta m_N$), and produces a class III TPE potential (100; 101).

3.3.1 EFT AND CHARGE DEPENDENT SCATTERING LENGTHS    Most of the work since since 1994 has been devoted to determining sources of $\Delta a_{CSB}$ other than the effects of $\rho^0 - \omega$ mixing. Coon and Niskanen (66) use a nonrelativistic formalism to derive a CSB two-pion exchange interaction that includes both CSB vertex corrections and the CSB effect of propagators arising from the different masses of the intermediate nucleon and $\Delta$ in different charge states. While the former are small, the latter gives a contribution of the sign and scale comparable to the experimental difference in the effective range parameters and the binding energy difference between $^3$H and $^3$He. In contrast, Li and Machleidt (102) use the Bonn meson-exchange model to make a systematic study of the CSB of the NN interaction caused by the nucleon mass splitting. Particular attention is paid to CSB generated by the $2\pi$-exchange contribution to the NN interaction, $\pi - \rho$ diagrams, and other multi-meson exchanges. The effects of the $\Delta$ are important in the first calculation but not in the second, yet both reproduce the observation. This indicates considerable model dependence in the details. However, the underlying origin of CSB in both calculations is the up-down quark mass difference; and the $nn$ force is more attractive than the $pp$ force in both calculations. Furthermore, both potentials account for the missing binding energy difference of A=3.

Li and Machleidt (103) also compute the charge-independence breaking (CIB) of the NN interaction due to pion-mass splitting. Besides the one-pion-exchange (OPE), they include the $2\pi$-exchange model and contributions from three and four irreducible pion exchanges. They find $\Delta a_{CD} = 4.7$ fm as compared with the empirical value of 5.7 ± 0.3 fm. If the effect (0.7 fm) of the $\pi\gamma$ exchange term (42; 46) is included, quantitative agreement is achieved.

A fit to NN phase shifts including various isospin-breaking interactions was carried out in Ref. (104), improving on an earlier analysis that used perturbative pions (105). The recent $\nu = 3$ work of Ref. (92) included charge dependence up to next-to-leading order of the counting scheme of Ref. (104). The effects of the pion mass difference in OPE and the Coulomb potential in $pp$ scattering is included, as is the pion mass difference in the leading-order TPEP, $\pi\gamma$ exchange (42), and two charge-dependent contact interactions. The empirical values of all

$^1S_0$ scattering lengths are reproduced. Similar results are obtained in Ref. (93).

The known charge-dependent effects of long and medium range account for the low-energy data. The only remaining unknown concerns the possible existence of very short-ranged, charge-dependent interactions. One might hope that calculations using non-relativistic quark and bag models to compute NN scattering might supply insight. Early calculations are reviewed in Ref. (30). There is a recent extensive review (106) and interesting later work (107; 108). The main motivation is to gain a better understanding of the short-ranged repulsion. Quark models use the quark Pauli principle and the gluon exchange hyperfine interaction to account for this. Then the CIB can be computed directly in terms of the mass difference between composite quarks and electromagnetic effects. The principal finding is that the quark-model effects responsible for the "hard" core of the NN interaction seem to harbor only small CIB effects.

Some of the successful quark models use the resonating group model in which the wave function is an antisymmetrized product of six single quark wave functions. This approach has been criticized by Miller (109) who argued that the inclusion of gluon degrees of freedom causes the exchange terms introduced by the antisymmetrizing operator to vanish. Furthermore, the direct use of quarks and gluons in non-perturbative calculations is still overly difficult.

3.3.2 CHARGE DEPENDENCE OF $\pi - N$ SCATTERING   Careful analysis (110) of low energy pion-nucleon scattering data give 7% CIB as measured by the "triangle discrepancy". The use of EFT (96) shows this to be a stunningly large effect. If one neglects electromagnetic and the kinematic effects of the charge-dependent masses, the leading-order chiral Lagrangian finds that the triangle discrepancy is given by:

$$D = \frac{T(\pi^+ p \to \pi^+ p) - T(\pi^- p \to \pi^- p) - \sqrt{2}T(\pi^- p \to \pi^0 n)}{T(\pi^+ p \to \pi^+ p) - T(\pi^- p \to \pi^- p) + \sqrt{2}T(\pi^- p \to \pi^0 n)} \approx \frac{\delta m_N - \bar{\delta} m_N}{4 m_\pi}. \tag{30}$$

Using $D = -7\%$ would lead to a huge **negative** value of $\delta m_N - \bar{\delta} m_N \approx -17$ MeV, indicating a serious problem with current understanding of low-energy QCD. However, electromagnetic effects are very important at low energies as is the kinematic difference in threshold energies. Kinematic effects and sub-leading strong interactions have been studied in Refs. (111) and (112), but much smaller effects are obtained. A better understanding of the electromagnetic corrections to the data is needed, and the data may be insufficiently precise to allow an accurate subtraction of separately-measured amplitudes.

## 3.4 EFT and Pion Production

Chiral symmetry restricts the form of pion-nucleon interactions, so understanding the production of a single pion in a NN collision is an important part of understanding QCD. As explained above, we expect the maximum breaking of charge symmetry to occur in $\pi^0$ interactions, so the production of neutral pions is our central topic.

The first step is to understand strong pion production. There is a long history (78) of phenomenological approaches that, while based on sound intuition and occasional success in reproducing data, lack a systematic expansion. It is reasonable to hope that using EFT would lead to a systematic understanding of pion production that could be applied to obtain insight about CSB.

Indeed recent analysis shows that it is possible to define a convergent effective field theory that allows a systematic study of the pion production amplitudes. There is a new challenge here. In a NN collision the initial momentum $p_i \sim \sqrt{m_\pi M_N}$ is larger than the pion mass by a factor of 2.6. This is still smaller than $M_{QCD}$, so the expansion should still converge, but more slowly. There are two options available to construct an EFT. The first, due to Weinberg, treats all light scales to be of order of $m_\pi$, so there is one expansion parameter: $\chi_W = m_\pi/M_N$. The other option is to expand in two scales simultaneously, namely $m_\pi$ and $p_i$. In this case the expansion parameter is

$$\chi = \sqrt{\frac{m_\pi}{M_N}} \sim 0.4 \; . \tag{31}$$

This scheme was originated in Refs. (113; 114) and applied in Ref. (115). The two additional scales, namely $\delta m$ and $m_\pi$, are identified with $\delta m/\Lambda_\chi \sim p_i/\Lambda_\chi = \chi$ and $m_\pi/\Lambda_\chi \sim p_i^2/\Lambda_\chi^2 = \chi^2$ where the former assignment is because of the numerical similarity of the two numbers ($\delta m = 2.1 \; m_\pi$, $p_i = 2.6 \; m_\pi$). Explicit calculations (78) were aimed at determining which option is better. In the Weinberg scheme some of the NNLO contributions exceed significantly the next–to–leading (NLO) terms. The chiral expansion converges only slowly, if at all. This point was further stressed in Ref. (116). As soon as the scale induced by the initial momentum is taken into account properly by expansion in $\chi$, the series indeed converges (114; 115) and certain divergent terms must be handled carefully (117). The latest $\chi$ expansion (118) led to a resolution of a long-standing puzzle. Including all of the next-to-leading order loop diagrams was shown to enhance the leading rescattering amplitude by a factor, 4/3, which seems to be just the value needed to understand the strong threshold cross section. The effects of these loop diagrams and higher order effects need to be included in pion production calculations.

We conclude this section by discussing the Lagrangian relevant for CSB $\pi^0$ production reactions. The strong interactions are the Weinberg-Tomozawa term and the standard axial-vector couplings – including recoil – of the pion to the nucleon and to the $\Delta$-isobar. The CSB interactions are encoded in Eq. (22)-Eq. (24), with various higher derivative terms included in Ref. (41).

## 4   Charge Symmetry Breaking in the Reaction $np \to d\pi^0$

Section 3.3 shows that EFT predicts substantial CSB in neutral pion interactions. There are few experimentally feasible measurements that test this prediction. It is therefore noteworthy that invariance under the charge symmetry Eq. (2), mandates that the deuteron angular distribution arising from the $np \to d\pi^0$ reaction is symmetric about 90° in the center of mass frame (cm). Fig. 1 illustrates this invariance. In going from the left to the right sides of Fig. 1, $P_{CS}$ interchanges neutrons and protons while the deuteron is unaffected. If charge symmetry holds there is no way to distinguish the left and right, so the deuteron yield at an angle $\theta$ must be equal to the yield at an angle $(\pi - \theta)$.

The size of CSB can be inferred from the cm forward–backward $np \to d\pi^0$ asymmetry, $A_{\text{fb}}$, defined as

$$A_{\text{fb}}(\theta) \equiv \frac{\sigma(\theta) - \sigma(\pi - \theta)}{\sigma(\theta) + \sigma(\pi - \theta)} \qquad A_{\text{fb}} \equiv \frac{\int_0^{\pi/2} d\Omega[\sigma(\theta) - \sigma(\pi - \theta)]}{\int_0^\pi d\Omega \sigma(\theta)} \tag{32}$$

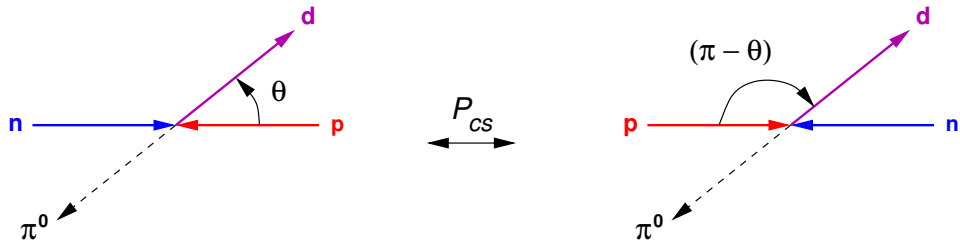

Figure 1: Charge symmetric relations for $np \to d\pi^0$ in the center of mass frame.

where $\theta$ is the cm angle between the incident neutron beam and the scattered deuteron. $A_{\text{fb}}$ has been examined at several energies (119; 120; 121; 122) as auxiliary results in experiments not specifically dedicated to its determination. The resulting lack of precision allowed only the determination of typical upper limts about 5-10 times larger than the signal eventually observed by Ref. (3).

In 1994 an experiment was proposed to measure $A_{\text{fb}}$ at an energy close to threshold using specific new techniques to eliminate or reduce systematic effects that could cause a false signal. At such energies the $np \to d\pi^0$ cm cross-section is given by

$$\frac{d\sigma}{d\Omega}(\theta) = A_0 + A_1 P_1(\cos\theta) + A_2 P_2(\cos\theta), \tag{33}$$

where $P_1$ and $P_2$ are Legendre polynomials. The $A_0$ and $A_2$ coefficients were previously measured (122) at a number of energies within 10 MeV above threshold. The presence of charge symmetry breaking is reflected in the presence of a non-vanishing $A_1 \cos\theta$ term. The angle-integrated form of $A_{\text{fb}}$ is $\frac{1}{2}A_1/A_0$. The results of this experiment were published in 2003 (3) and are discussed here.

## 4.1 Experiment

Detecting an asymmetry in the produced neutral pions directly would involve the essentially impossible challenge of detecting an asymmetry in the decay photons. Therefore deuterons were detected.

The experiment was performed at TRIUMF using a 279.5-MeV neutron beam, a liquid hydrogen target, and the SASP magnetic spectrometer (123) positioned at 0°. This reaction energy is a few MeV over the 275.06-MeV reaction threshold, causing the deuterons from $np \to d\pi^0$ to form a small forward cone centered about the neutron beam direction and a distinct kinematic locus in momentum versus laboratory scattering angle, which is shown in Fig. 2 for the collected data. The background surrounding the locus is from $(n, d)$ reactions on nuclei such as carbon in materials near the target. The deuterons with higher (lower) momentum were produced in the forward (backward) direction in the center of mass frame so that comparing the top half of the locus with the bottom half is an indication of $A_{\text{fb}}$. Due to the effects of multiple scattering, energy loss, and other physical processes, extracting $A_{\text{fb}}$ directly from the data was impossible. Consequently, the angle integrated $A_{\text{fb}}$ was determined by comparing the data with a carefully calibrated simulation of the experiment.

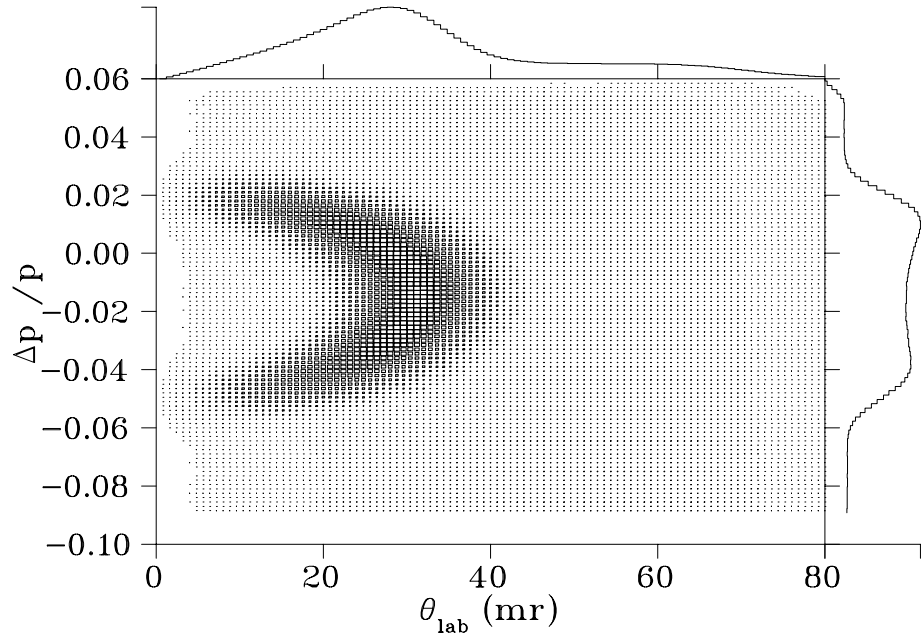

Figure 2: Percent momentum versus lab scattering angle of the detected deuterons. The kinematics of $np \to d\pi^0$ at threshold constrain deuterons from that reaction into the dark locus; deuterons from $(n,d)$ reactions on other nuclei near the target populate the rest of the plot. The curve above the 2-D plot shows the lab scattering angle of the detected deuterons and peaks at $\sim 28$ mrad. The curve to the right of the 2-D plot is the percent momentum, $\delta = \Delta p/p$, of the detected deuterons. Reprinted figure with permission from A.K. Opper, *et al.*, Phys. Rev. Lett. **91** 212302 (2003). Copyright 2003 by the American Physical Society.

In measuring an asymmetry, such as $A_{\text{fb}}$, conditions that might systematically affect the foward-going deuterons differently from the backward-going deuterons would ideally be eliminated from the experimental setup or "flipped" to cancel their effects. This possibility was exploited to the extent that all the deuterons were detected in one single magnetic setting of the spectrometer at 0° so that the material traversed, the luminosity and live time for the forward- and backward-going deuterons were the same. Systematic effects which could not be cancelled in this way, such as focal plane detector efficiencies, were measured and reproduced in the simulation.

A schematic drawing of the experimental apparatus is shown in Fig. 3. The TRIUMF CHARGEX facility (124) produced a neutron beam at 0° by passing a high intensity proton beam, typically 350 nA, through a 220 mg/cm$^2$, 7 mm high $^7$Li target and initiating the $^7$Li(p,n)$^7$Be reaction. The momentum of the primary proton beam was vertically dispersed and the $^7$Li target relatively short (vertically) so as to minimize the momentum range of the primary beam which produced the neutron beam. A sweeping magnet deflected the primary proton beam away from the experimental area. The liquid hydrogen target (LH$_2$) was centered 92 cm downstream from the $^7$Li target and was contained within a flat cylindrical volume 10 cm in diameter with a nominal thickness of 2 cm. Two sets of veto counters (FEV1, FEV2) and a trigger counter set (FET) were each composed of a pair of plastic scintillators positioned above one another. This allowed more stable operation in the high (few MHz) particle rate environment. The thick veto scintillators were upstream of the LH$_2$ and shadowed it. The FET counters were positioned immediately downstream of the LH$_2$. The FET counter thresholds were set to discriminate against protons but trigger on the higher pulse height deuterons. For calibration runs, such as $(n, p)$ elastic runs, the FET thresholds were set for protons. The FEV thresholds were always set for protons. The output of both halves of each of the FET, FEV1, and FEV2 counters were scaled and then combined in a logic "OR" to form FET, FEV1, and FEV2 signals. The front-end trigger for $np \to d\pi^0$ was $\overline{\text{FEV2}} \cdot \text{FET}$ to ensure that an uncharged particle entered the target and a charged particle exited it.

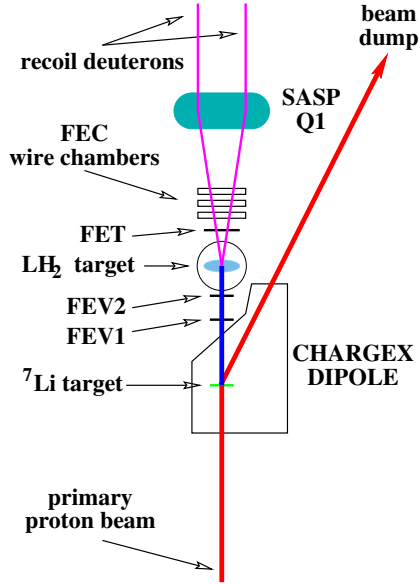

Figure 3: Schematic of the $np \to d\pi^0$ experimental apparatus

Three multi-wire proportional chambers, positioned upstream of the SASP entrance (FECs *i.e.* Front-End Chambers), provided tracking information for charged particles. Each FEC consisted of a pair of orthogonal wire planes. The first and last FECs, separated by 33 cm, were mounted to measure vertical and horizontal coordinates. The third FEC was positioned midway between the other two and rotated 40° with respect to them for efficiency measurements and to aid in multi-hit track reconstruction. Each FEC wireplane typically ran at greater than 98% efficiency.

The SASP spectrometer is a vertical bend QQD spectrometer with a nominal angular acceptance of 103 mr × 42 mr = 13.5 msr and a momentum acceptance of $\delta = (p - p_0)/p_0 = (-10\%, +15\%)$, where $p$ is the momentum of the particle moving through SASP and $p_0$ is the central momentum. Particle tracking near the SASP focal plane was provided by two vertical-drift chambers (VDCs). Each chamber contained an $X$ plane to measure position in the bend direction and a "U" plane inclined at 30° to $X$ from which the non-bend coordinate could be computed. The first VDC lay parallel to the focal plane and 3 cm above it, and the second VDC was an additional 56 cm above the first. Three sets of scintillators, downstream from the VDCs, provided timing and particle ID information as well as sufficient redundancy to determine the efficiencies of all focal plane area detectors. Measurements of $np$ elastic scattering with incident neutron beams that filled the same target space and produced protons that spanned the momentum distribution of the $np \to d\pi^0$ reaction provided a stringent test of the description of the spectrometer acceptance. Further details on the apparatus and other technical aspects of the measurement are found in Ref. (125).

The angular distribution of the $np \to d\pi^0$ data itself was used to determine the mean neutron beam direction. The data were first analyzed with nominal values of parameters describing the incident beam trajectory and position at the $^7$Li target. A two-dimensional histogram of the horizontal and vertical components of the reaction angle was produced for events within a small range of focal plane momenta in order to select deuterons emitted near 90° in the center-of-mass system. The resulting "smoke ring" pattern represented the azimuthal distribution of the $np \to d\pi^0$ events with the assumed incident beam trajectory. The offset of the center of this pattern from zero in each of the horizontal and vertical directions (i.e. in the horizontal and vertical projections of the reaction angle) was used to adjust the parameters describing the incident beam direction.

The simulation, based on GEANT3, a well-known simulation tool that describes the passage of elementary particles through matter, begins with a proton beam incident on the $^7$Li target and includes energy loss by the proton beam as well as the angular and energy distribution of neutrons from the $^7$Li(p, n) reaction. Production of deuterons according to the distribution of Eq. (33) is allowed in the LH$_2$ target and other hydrogenous material such as scintillators and their wrapping. Deuteron energy loss and multiple scattering are handled using standard GEANT tracking options. However, deuteron reactions which amount to 1–2% losses in detector materials and are significantly momentum dependent over the 8% momentum range of the experiment, had to be determined and included in the simulation. Some fraction of the reaction cross section is "gentle breakup" of the deuteron, which leaves a neutron and proton with roughly the same speed and direction as the original deuteron. Tables of total reaction cross section vs. incident beam energy for d +$^{12}$C and for d +$^{16}$O (126) were assembled and fit to simulate the reaction losses. Tracking through the SASP dipole used a field map obtained at 875 A and scaled up to the operating current of 905 A. Data were acquired in 10 different periods spanning two years and the simulation accounted for measured detector efficiencies, scintillator thresholds, missing FEC wires, and known changes in target thickness in a manner consistent with the actual running periods. A blind analysis technique was used that incorporated a hidden offset to the $A_1/A_0$ asymmetry parameter of the $np \to d\pi^0$ generator. The collaborators developing the simulation and extracting the observable did not know the value of the offset until all consistency checks had been satisfied.

Robust tests of the simulation involved comparing the model and data for observables that are independent of $A_{\text{fb}}$. To that end, two sets of acceptance calibration data were obtained: (1) measurements of the SASP dispersion, taken with high resolution $^{208}$Pb(p,p') (small beamspot on target), and (2) measurements of $(n,p)$ elastic scattering (large secondary beam on target). The $^{208}$Pb(p,p') calibration data were taken with constant fields on the SASP magnets and different beam energies to span the momentum range of interest. Extracting the dependence of the projectile momentum on its detected position in the focal plane detectors ($X_F$) from these data led to a third order polynomial of $X_F$. The measurements of $np$ elastic scattering were made with the SASP magnets set to their nominal values, and the primary beam energy adjusted so that the elastically scattered protons had a momentum deviation $\delta = -4, 0$ or $+4\%$ compared to the central momentum of the deuterons of interest. The acceptance of the SASP is a function of the initial position and direction of the deuteron and its momentum ($X_i$, $\theta_i$, $Y_i$, $\phi_i$, and $\delta$). Describing this acceptance properly required an accurate model of the magnetic fields and interior surfaces of the SASP because deuterons could collide with interior surfaces of the SASP and be lost from the locus and the reconstructed momentum variable, which is a function of the focal plane variables $X_f$ and $\theta_f$, may have been distorted in a momentum dependent way by an inaccurate model. Non–uniformities in the momentum acceptance of the SASP would systematically produce a false asymmetry and had to be limited.

High-statistics data from $np$ elastic scattering were collected and compared to model simulations to determine a fiducial volume of uniform acceptance. The $np$ elastic scattering data provided the best means to determine the acceptance because the incident beam of this reaction filled the target parameter space in a manner similar to that of the $np \to d\pi^0$ reaction. The momentum distributions of the deuterons from $np \to d\pi^0$ and of elastically scattered protons at the different beam energies are shown in Fig. 4.

To investigate and eliminate any momentum dependent effects which could potentially result in false $np \to d\pi^0$ forward-backward asymmetries, projections of $\theta_i$ and $\phi_i$ were made for slices in $X_i$ and $Y_i$, respectively, and ratios of these distributions were made for the $-4\%$ and $+4\%$ momentum sets for both data and simulation. The portions of these ratio distributions that were flat (i.e. constant over position, angle, and $\delta$) and common to both data and simulation defined the fiducial acceptance of the spectrometer.

There are two distinct ways in which parameters affect counts per bin in $\delta - \theta_{lab}$ space. In one case a change in parameter alters a pixel and all its immediate neighbors in the same sense and by very nearly the same percentage. The parameter $A_1/A_0$, momentum dependent deuteron reaction losses, and the SASP acceptance are in this category. The fiducial acceptance, defined by the cuts described above, limits the effects from any momentum dependence in reaction losses or acceptance, because only regions with no momentum dependence are within the cuts. The other way parameter changes affect bin counts is when the locus position shifts so that the change in number of counts depends on the local gradient in counts per pixel. Such parameters include the $LH_2$ target thickness, proton beam energy, and SASP central momentum. Model defects affecting the profile (width, extent of tail) or position of the locus have greater potential for skewing the fit of parameters of the second type (locus shifters) than for skewing the first type (fixed-locus).

Comparisons among simulations were carried out to determine how experimental parameters and other effects were correlated with $A_1/A_0$. For example, momentum dependent deuteron reaction losses and detection efficiencies are obvious mechanisms which can mimic the effect of a non–zero $A_1/A_0$. Combining each correlation with the independently-determined uncertainty of its parameter gave the systematic contributions to $A_1/A_0$, which when added in quadrature gave a total systematic uncertainty of $5.5 \times 10^{-4}$. For the LH$_2$ target thickness, the proton beam energy ($T_{\text{beam}}$), and the central momentum of the SASP ($p_0$), the independent information was not a sufficient constraint. Therefore, these three parameters, along with $A_1/A_0$, were treated as free parameters and their values extracted from fitting the data. Simulations were made and $\chi^2$ calculated for 81 points in a four-dimensional space in which each of the four free parameters was stepped above and below a nominal value. $\chi^2$ minimization techniques (127) were used to obtain the values of the parameters at the global minimum, while the local curvature of the surface gave their errors and mutual correlations.

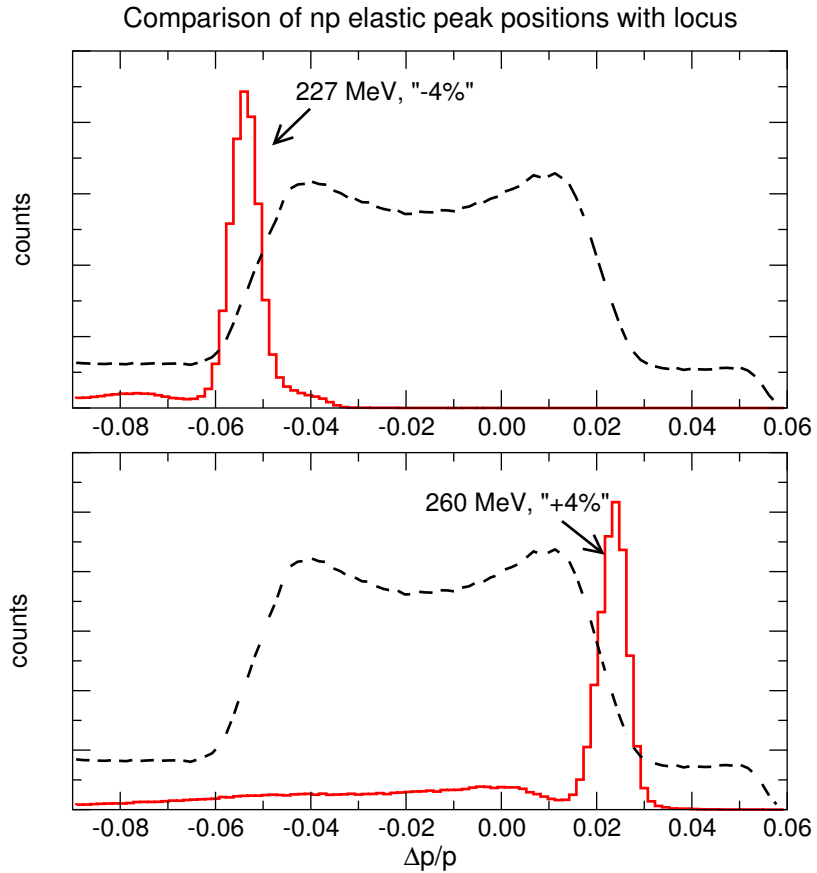

Figure 4: Momentum projection of deuterons from $np \to d\pi^0$ (dashed line) with momentum of elastically scattered protons at different incident energies overlayed (solid line).

Further rebinning was done to remove sensitivity to unimportant details of the simulation without losing sensitivity to $A_1/A_0$. The $\chi^2$ grid search was repeated using 20 bins in $\delta$ and 10 bins in $\theta$, and again with 10 bins in $\delta$ and 5 bins in $\theta$. As expected, the fractional error dropped to 2.1% and 1.4%, respectively, with $A_1/A_0$ remaining consistent within errors. A more sophisticated binning scheme which treated the locus as a set of "elliptical" and "radial" bins on top of rectangular background bins produced an rms error of 2.1% for 2500 background bins and 36 locus bins, and an rms error of 1.2% for 100 background bins and

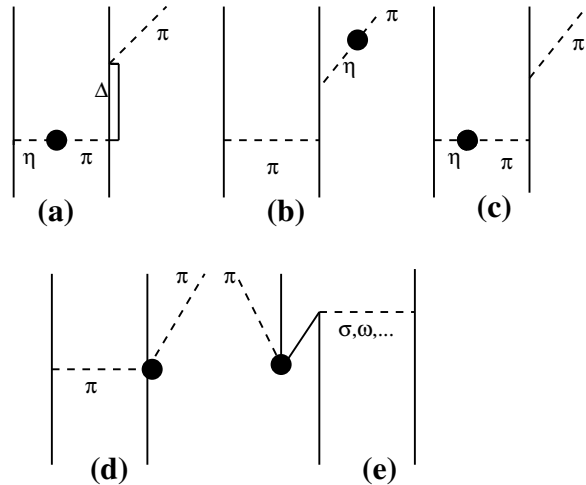

Figure 5: $\pi^0$ production diagrams; the circle indicates a CSB mechanism

6 locus bins. In all fits and binning schemes the best fit values of $A_1/A_0$ in all acceptance subspaces agreed within errors with the value for the full acceptance, which is $(34.4 \pm 16) \times 10^{-4}$. This suggests that any remaining systematic effects from deficiencies in the acceptance model are within errors. The result of the experiment is that

$$A_{\text{fb}} = [17.2 \pm 8(\text{stat}) \pm 5.5(\text{sys})] \times 10^{-4}. \tag{34}$$

## 4.2 Interpretation

The first calculations of $A_{\text{fb}}$ (45) obtained a large negative value at 577 MeV, using a mechanism involving an intermediate $\Delta$ as seen in Fig. 5a. Later calculations by Niskanen and collaborators (128; 129) included the additional sizeable effects of $\rho^0 - \omega$ mixing (which depend on the unmeasured $\rho N\Delta$ coupling constant) and the effects of CSB direct emission of the pion driven by $\pi^0 - \eta$ mixing as shown in Fig. 5b,c.

The first calculation at threshold was made by Niskanen (130) using a meson-exchange coupled-channel model which showed a major contribution due to $\pi^0 - \eta$ mixing in both the exchanged and produced (outgoing) meson, as illustrated in Fig. 5a-c. At the energy of this measurement the prediction was $A_{\text{fb}} = -28 \times 10^{-4}$ using an $\eta NN$ coupling constant of $g^2_{\eta NN}/4\pi = 3.68$ from meson exchange NN potential models (36) and a $\pi^0 - \eta$ mixing matrix element $\langle \pi^0|\mathcal{H}|\eta\rangle = -5900$ MeV$^2$ (131). Using the upper limit of Eq. (9) and the more recently determined mixing matrix of $-4200$ MeV$^2$ (34) reduces the magnitude of this prediction to $A_{\text{fb}} = -3.2 \times 10^{-4}$. The effects of $\rho^0 - \omega$ mixing were not included.

Inspired by the maximal breaking of charge symmetry in $\pi^0 - N$ interactions implied by Eq. (22) and Eq. (23), Ref. (96) used chiral EFT to compute $A_{\text{fb}}$. Fig. 5d displays these CSB seagull terms. These terms yield positive contributions to $A_{\text{fb}}$.

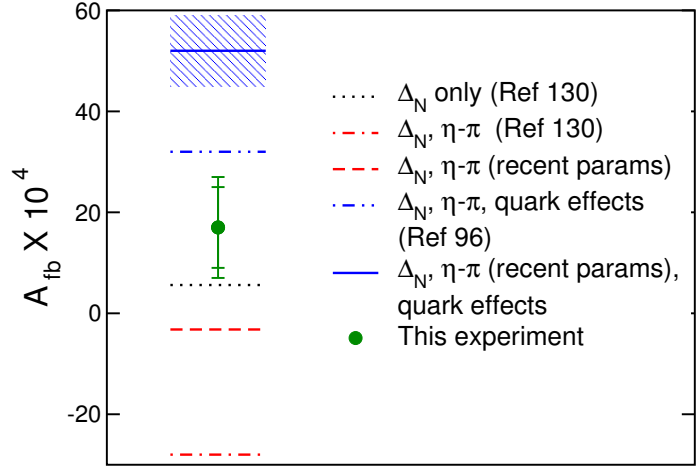

Figure 6: Measured value of $A_{\text{fb}}$, at incident $T_{\text{neutron}} = 279.5$ MeV, and the published results of Refs. (96; 130). The dots represent the effects of the $n - p$ mass difference ($\Delta_N$) in the emission vertex. Also including $\pi^0 - \eta$ mixing leads to the dashed line, and further including the quark effects of Eq. (22) and Eq. (23) leads to the solid line.

The non-vanishing $A_{\text{fb}}$ arises from interference between strong and CSB amplitudes, but neither of the near-threshold calculations (96; 130) obtained very good agreement with experiment, and both were made prior to the current understanding of convergence (118). There is also great sensitivity to the specific values of input parameters. Despite these caveats, it is worthwhile to analyze the experiment using these currently available tools. At the energy of the experiment, $A_{\text{fb}}$ was estimated (based on Ref. (96)) to be

$$A_{\text{fb}} = \frac{5.6}{10^4} - \frac{8.8}{10^4} \frac{g_{\eta NN}}{\sqrt{4\pi \cdot 0.51}} \frac{\langle \pi^0 | \mathcal{H} | \eta \rangle}{-4200 \text{ MeV}^2} + \frac{50.0}{10^4} \frac{(\delta m_N - \frac{1}{2}\bar{\delta} m_N)}{2.4 \text{ MeV}}, \quad (35)$$

where the first term $\Delta - N$ arises from the $n - p$ mass difference, the second from $\pi^0 - \eta$ mixing, and the third new term from $\pi^0 - N$ scattering. Including the new term causes $A_{\text{fb}}$ to be positive instead of negative. If one uses the value and error from Eq. (25) and the smaller value of Eq. (9), $g_{\eta NN}[= \sqrt{4\pi \cdot 0.10}]$, then $A_{\text{fb}} = +(52 \pm 7) \times 10^{-4}$, close to the experimental value of Eq. (34) (but larger than the value $32 \times 10^{-4}$ given in Ref. (96)). The agreement is as good as one can expect now because the effects of the loop correction, which double the size of the computed strong cross section (118), are not included. Figure 6 shows the experimentally determined value of Eq. (34) and the published results of Refs. (96; 130).

The measured positive value of $A_{\text{fb}}$ suggests that the effects of charge-symmetry violating $\pi^0 - N$ interactions as outlined in Ref. (96) and driven by the constraints of chiral symmetry have been observed. A theory collaboration CSB theory has been formed (41; 132) to study CSB in $\pi^0$ production reactions. Recent progress in achieving a convergent power counting in the $NN \to NN\pi^+$ reaction (118) and obtaining agreement with $pp \to d\pi^+$ data allows hope that a quantitative understanding of $A_{\rm fb}$ will be possible in the not-too-distant future.

## 5 Observation of the Reaction $dd \to \alpha\pi^0$

The reaction $dd \to \alpha\pi^0$ is forbidden if isospin is conserved because the isospin of the pion is unity while that of the other particles is zero. This reaction also tests charge symmetry as defined by Eq. (3) because the $\pi^0$ wave function is odd under $P_{\rm cs}$ while the deuteron and $\alpha$-particle are even (self-conjugate) under charge symmetry, or $P_{\rm cs}|d\rangle = |d\rangle$, $P_{\rm cs}|\alpha\rangle = |\alpha\rangle$, and $P_{\rm cs}|\pi^0\rangle = -|\pi^0\rangle$. Thus $\langle dd|H_S|\alpha\pi^0\rangle = \langle dd|P_{\rm cs}^\dagger H_S P_{\rm cs}|\alpha\pi^0\rangle = -\langle dd|H_S|\alpha\pi^0\rangle = 0$ if Eq. (3) holds. A violation of Eq. (1), that is not a violation of Eq. (3), could occur with the existence of a non-vanishing commutator of $H_S$ that is an iso-tensor operator of rank two or greater. This could happen if the left and right hand sides of a $\pi^0$ production reaction involving self-conjugate initial and final states differed by *two* units of isospin.

To observe this reaction is to observe the square of a CSB amplitude, so achieving a successful measurement involves meeting an enormous technical challenge. To be explicit, we make a simple estimate of the cross section by starting with the total cross section for the CS allowed $dp \to {}^3{\rm He}\,\pi^0$ reaction (133), including the suppressive effects of the spectator neutron (134), and multiplying the amplitude by a suitable CSB scaling factor, $(m_d - m_u)/m_N \sim 1/300$. This results in an estimated cross section of about 10-20 pb.

### 5.1 Experiment

There have been several attempts to observe the $dd \to \alpha\pi^0$ reaction (see the review by Banaigs (135)) since it was posed as a test of charge symmetry in 1956 (136). In 1991, a group at Saturne reported $\alpha\gamma$ coincidences at a rate of $0.97 \pm 0.20 \pm 0.15$ pb/sr (137) using 1.10 GeV deuterons. Later Dobrokhotov observed that the double radiative capture and charge symmetry allowed process $dd \to \alpha\gamma\gamma$ could have produced such coincidences at the reported rate (138), so the Saturne result is not evidence for the observation of the CSB $dd \to \alpha\pi^0$ reaction. This made it crucially important for any subsequent search to clearly separate the $\alpha\pi^0$ channel from the double radiative capture continuum.

When the IUCF group designed their search for the $dd \to \alpha\pi^0$ reaction using the Cooler storage ring, two improvements were made. First, both $\pi^0$ decay photons were detected in a Pb-glass Čerenkov array that was insensitive to other reaction products, thus ensuring a nearly background-free set of candidate events. Second, the forward-going $\alpha$-particles produced just above threshold were separated magnetically from the deuteron beam and sent into a channel that precisely measured their scattering angle and momentum. This made it possible to reconstruct the $\pi^0$ missing mass from $\alpha$-particle kinematics alone. (Energy and angle information from the Pb-glass array was too poor to make an effective reconstruction.)

The experimental layout at IUCF is shown in Fig. 7. The gas jet target, located between the two Pb-glass arrays, was fed through a nozzle cooled to 40 K. This windowless target provided a source of events essentially free of contamination from pions generated by deuteron bombardment of heavier nuclei. Even so, there was a considerable flux of $\alpha$-particles traveling with the beam from $(d, \alpha)$ reactions on residual gas and beam line hardware, making the coincidence with photons in the Pb-glass array essential for identifying candidate $dd \to \alpha\pi^0$ events. Differential pumping in the target region limited the placement of Pb-glass modules to two large arrays on the left and right sides of the beam. Monte Carlo simulations as well as $pd \to {}^3\text{He}\ \pi^0$ calibrations of this arrangement gave an efficiency for the capture of both $\pi^0$ photons of $36 \pm 1\%$.

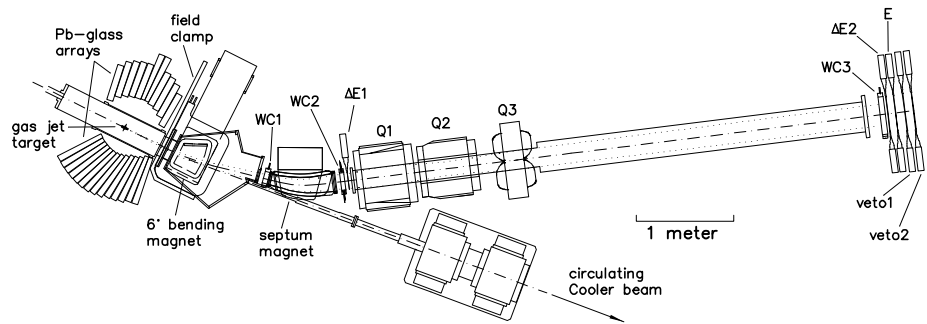

Figure 7: A layout of the IUCF detectors. Components are labeled, including wire chambers (WCn), plastic scintillation detectors ($\Delta$En, E, and Veto-n), and quadrupole magnets (Qn). Reprinted figure with permission from E.J. Stephenson et al., Phys. Rev. Lett. **91**, 142302 (2003). Copyright 2003 by the American Physical Society.

An important aspect of the IUCF design was the magnetic channel for recoil $\alpha$-particles. The first magnet after the jet target (a part of the Cooler ring) bent the beam by 6° and the $\alpha$-particles by 12.5°. The requirement to collect all of the forward cone of recoil $\alpha$-particles in the next 20° septum magnet limited the deuteron bombarding energy to less than about 231 MeV, just 5.5 MeV above the $\pi^0$ production threshold. In order to use the recoil $\alpha$-particles for an accurate reconstruction of the $\pi^0$ mass, it was necessary to measure both the scattering angle and the $\alpha$ momentum. The laboratory scattering angle was determined by the first multiwire chamber (WC1) located just in front of the septum magnet. The momentum information was calculated from a time-of-flight measurement between the $\Delta$E1 and $\Delta$E2 scintillators. But energy loss in the scintillators and wire chambers slowed the $\alpha$-particles, requiring the momentum to be determined from a model that included the $\alpha$ energy loss. During the course of the experiment, it was also necessary to track and correct the time of flight for changes in the time offset introduced by each photomultiplier tube and its associated electronics. Scintillator pulse height information was used to identify the $\alpha$-particles. The scintillator coincidence trigger for the channel was adjusted so that its time acceptance window would eliminate the bulk of the protons arising from deuteron breakup and going through the magnetic channel.

Figure 8 illustrates that after a simple set of selection criteria were met, the set of candidate events contained only $\alpha$-particles and two energetic photons and thus must consist of examples of either the $dd \to \alpha\pi^0$ or the $dd \to \alpha\gamma\gamma$ reaction. In both cases the opening angle between the two photons was close to 180°. For double radiative capture, this arose because the magnetic channel selected $\alpha$-particles corresponding to a missing mass near the $\pi^0$ mass. The left panel in Fig. 8 shows the time of photon arrival relative to the $\alpha$-particle time versus the recorded energy for a photon in the beam right array. With the requirement that only a forward $\alpha$-particle be detected in the magnetic channel, there is in this spectrum both a localized group of higher energy photons and a larger group of lower energy coincidence events spread randomly in time. The box in the left panel surrounds those photons chosen as candidates at this stage of the analysis. The right panel shows the signal that appeared in the beam left Pb-glass array in coincidence with an $\alpha$-particle and a beam right photon (inside the left panel box). In this case, there is a clear and clean group of higher energy photons. With a similar choice of analysis box, essentially all random photons were excluded.

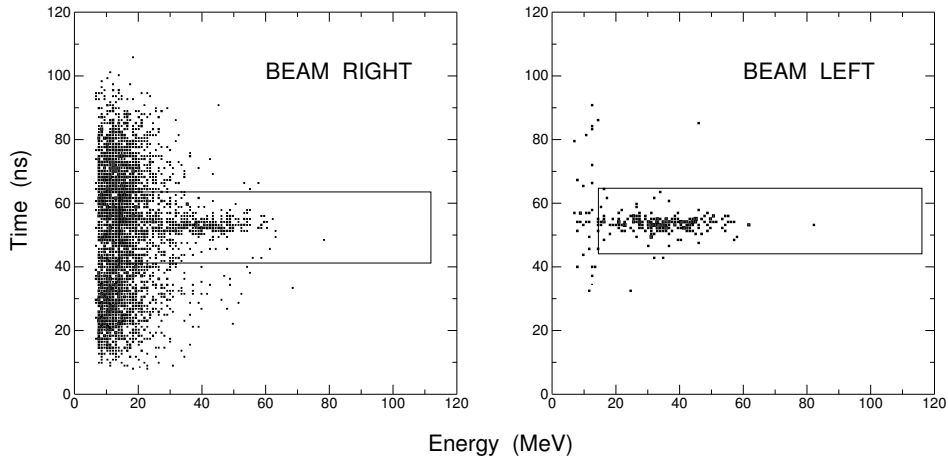

Figure 8: Scatter plots of the time of arrival of a photon in the Pb-glass array relative to the arrival of the $\alpha$-particle in the magnetic channel versus the recorded energy of the photon. The box within each panel shows the selection gate for good events. The left panel contains photons that were recorded in the beam right aray and gated only on the presence of an $\alpha$-particle in the channel. The right panel shows the events in the beam left Pb-glass array gated on both the $\alpha$-particle and the presence of a photon within the box gate on the other array (left panel).

The IUCF experiment ran for two months during the summer of 2002. It started at 228.5 MeV, an energy chosen to ensure that all $\alpha$-particles were inside the magnetic channel acceptance. After one month, preliminary online analysis of the candidate events (gated on an $\alpha$-particle in the magnetic channel and two photons, one in each half of the Pb-glass array) yielded a version of the left spectrum in Fig. 9 with poorer mass resolution. (The mass resolution was improved offline by adjusting the photomultiplier tube timing offsets.) At that time, the decision was made to increase the deuteron bombarding energy to 231.8 MeV, even if some of events (eventually estimated by Monte Carlo to be 10%) were lost at the edge of the channel. When the peak remained at the $\pi^0$ mass, its identification as the CSB $dd \to \alpha\pi^0$ reaction was confirmed. At the end of the analysis, both peaks reproduced the $\pi^0$ mass to within their statistical precision of 60 keV. Peak sums at the two energies of $66 \pm 11$ and $50 \pm 10$ events were obtained from a fit to the spectra using a Gaussian shape for the $\pi^0$ mass peak. The shape of the $\alpha\gamma\gamma$ continuum, but not its magnitude, was based on the expected double radiative capture continuum shape folded with the acceptance of the magnetic channel and plotted as a function of missing mass.

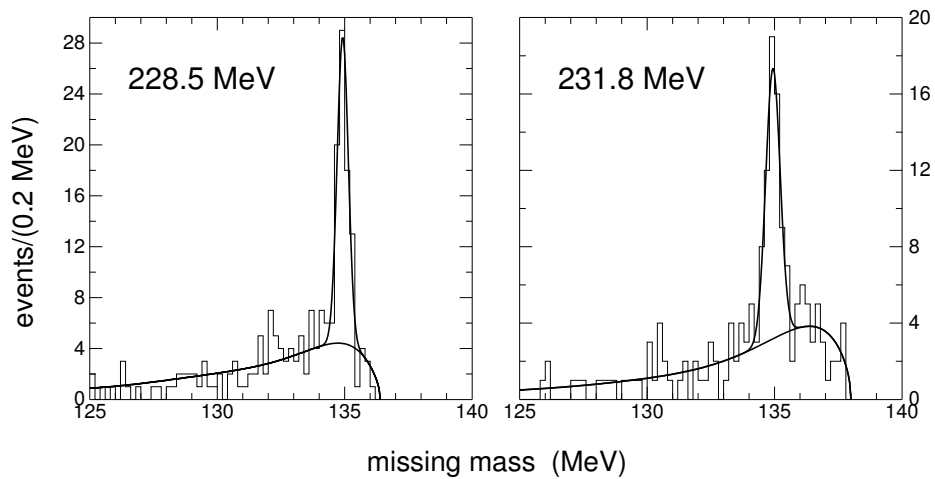

Figure 9: Missing mass histograms of candidate events at the two energies of the IUCF experiment. The curves show the best fit to the data using a Gaussian shape for the $\pi^0$ peak and an empirical shape (with only an adjustable overall scale) for the double radiative capture continuum. Reprinted figure with permission from E.J. Stephenson *et al.*, Phys. Rev. Lett. **91**, 142302 (2003). Copyright 2003 by the American Physical Society.

The luminosity (product of target thickness and beam current) for an internal target experiment at the IUCF Cooler is best obtained by scaling to some monitor reaction running in parallel with the experiment. For this case, $d + d$ elastic scattering at $\theta_{c.m.} = 90°$ (as observed by the detectors at 44° in Fig. 10) served as the online monitor. This cross section was calibrated against $d + p$ scattering by using HD molecular gas as the target and adding two scintillator telescopes mounted to cover the scattering angles near 25°. To make corrections for the distribution of the target gas along the beam line, a position-sensitive silicon detector was mounted to observe the recoils for $d + d$ elastic scattering in coincidence with a forward scintillator (not shown). The acceptance of the detectors was calculated using a Monte Carlo simulation. The $d + p$ reference cross sections were taken from a polynomial fit to the data of Ermisch (139). This gave $dd \to \alpha\pi^0$ total cross sections of $12.7 \pm 2.2$ pb and $15.1 \pm 3.1$ pb for the two energies of 228.5 and 231.8 MeV. Recently, the accuracy of the Ermisch cross sections has been challenged by new measurements from RIKEN and RCNP that are some 30% lower at 135 MeV (140). This discrepancy is not resolved; new measurements over a range of energies are needed.

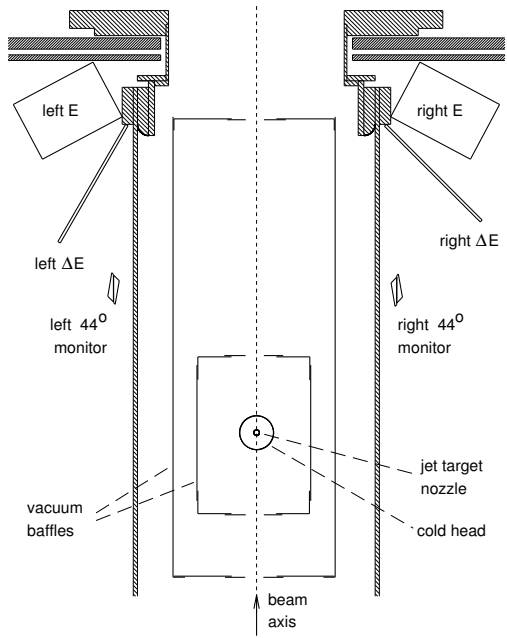

Figure 10: Layout of the monitor and cross calibration scintillators used to normalize the cross section for the $dd \to \alpha\pi^0$ reaction. The 44° detectors observed $d + d$ elastic scattering coincidences. Each also operated independently in coincidence with an opposite side 25° system to observe $d + p$ elastic scattering (using an HD target). Other structures include vacuum pumping baffles and beam line hardware. The left and right $\Delta$E-E telescopes were removed whenever the Pb-glass arrays were rolled into place.

The total cross sections at the two energies are consistent with S-wave pion production with an average cross section of $\sigma_{\rm TOT}/\eta = 80 \pm 11$ pb where $\eta = p_\pi/m_\pi$. These results produced the first observation of the $dd \to \alpha\pi^0$ CSB reaction as well as the related $dd \to \alpha\gamma\gamma$ double radiative capture process. While corrections to the cross section for channel and Pb-glass efficiency are significant, the systematic errors of 6.6% remain less than the statistical errors in this experiment.

## 5.2 Interpretation

The CSB amplitudes for the recently observed $dd \to \alpha\pi^0$ reaction were investigated (41) using the $\chi$ counting scheme [Eq. (31)] of chiral perturbation theory (113; 115) to classify and identify the leading-order terms.

At leading order (LO) there is only one contribution: pion rescattering in which the CSB occurs through the seagull pion-nucleon terms linked to the nucleon mass splitting in Eq. (22) and Eq. (23), as seen in Fig. 5d. The irreducible part of this diagram is $\mathcal{O}[\epsilon m_\pi^2/(f_\pi^3 M_p)]$.

There is no NLO contribution, but there are several NNLO contributions, especially a term in which a one-body CSB operator ($\propto \beta_1 + \bar{\beta}_3$) is sandwiched between initial and final state wave functions, as illustrated by the example of Fig. 5b. Photon exchange in the initial state also enters at this order, but is not included in the calculation. Loop diagrams enter at NNLO. Another set of terms that arises at order N4LO involve EFT LECs. These were estimated as seen in Fig. 5e using $\pi^0 - \eta$ mixing to generate CSB and resonance saturation using the exchange of heavy mesons along with a Z-graph mechanism to estimate the influence of short-range physics. Two other CSB Z-graphs arise from the exchange of a photon and the exchange of a mixed $\rho^0 - \omega$.

As a first step, a few tree-level diagrams were evaluated (41) using a plane-wave approximation for the initial $dd$ state, treating the deuteron as a $^3S_1$ state with a Gaussian relative wave function and modelling the $\alpha$ wave function as a Gaussian with relative s-waves. With these major simplifications the leading order pion exchange term (which is even in isospin and odd in spin operators) vanishes. Nevertheless, evaluating the higher order terms (obtained with the larger value of $g_{\eta NN}$ of Eq. (9)) yields cross sections of 23 and 31 pb, to be compared with the $12.7 \pm 2.2$ and $15.1 \pm 3.1$ pb measured at the two IUCF energies. If instead the smaller value of $g_{\eta NN}$ is used the results would be 13 pb and 17 pb, in accidental agreement with the observation.

It is interesting to examine the computed relative proportions of the pion exchange, photon exchange (Z-graph term), exchange of a mixed $(\rho^0 - \omega)$, and exchange of a mixed $(\pi^0 - \eta)$ (sum of one-body and heavy meson exchange between nucleons) contributions to the matrix element. These are roughly $\pi:\gamma:(\rho^0-\omega):(\pi^0-\eta)=1:11:12:21$. The total cross section may be expressed in terms of the relative contributions of the different CSB mechanisms with the dependences on the corresponding parameters made explicit:

$$\sigma(228.5 \text{ MeV}) = (23.0 \text{ pb}) \left( 0.254 + 0.0188 \frac{\delta m_N}{2.03 \text{ MeV}} + 0.0034 \frac{\bar{\delta} m_N}{-0.74 \text{ MeV}} \right.$$
$$\left. + 0.456 \frac{g_{\eta NN}}{\sqrt{4\pi \cdot 0.51}} \frac{\langle \eta | H | \pi^0 \rangle}{(-4200 \text{ MeV}^2)} + 0.268 \frac{g_{\rho NN} g_{\omega NN}}{4\pi \sqrt{0.43 \cdot 10.6}} \frac{\langle \omega | H | \rho^0 \rangle}{(-4300 \text{ MeV}^2)} \right)^2 . \quad (36)$$

The numerical coefficients are the fractions of the transition matrix element corresponding to each mechanism, using the parameters (including the positive relative signs of all of the meson-nucleon coupling constants) of Ref. (41). The purely numerical term is due to one-photon exchange Z-graph. This estimate is dominated by the influence of the Z-graphs as 62% of the $\pi^0 - \eta$ term arises from a Z-graph. So do all of the other terms, except for the very small pion exchange terms that arise mainly from a higher-order recoil correction. Indeed, eliminating the Z-graphs would reduce the estimated cross section to 0.9 pb. The large size of the Z-graphs is not in accord with estimates based on power counting. A general remark is that the Z-graph terms need to be computed using the same interactions that generate the initial scattering and final bound state wave functions.

The CSB theory collaboration (41; 132) has recently used realistic two- and three-nculeon interactions (88; 142) to obtain bound state wave functions with significant high momentum components. This increases the computed size of the one body term by about a factor of 8.

The effects of initial state interactions obtained using the same potentials were also included (141). These spin and isospin dependent interactions change the influence of the specific spin-isospin dependence that causes the pion exchange term to be negligible.

The results of this newer calculation may be parametrized as

$$T = \frac{\delta m_N}{2.05 \text{ MeV}} T_\pi + \frac{\bar{\delta} m_N}{-0.76 \text{ MeV}} \bar{T}_\pi + \mathcal{T} \qquad (37)$$

$$\mathcal{T} = \tilde{\beta}(T_{OB} + \frac{g^2_{\sigma NN}}{4\pi \cdot 7.1} T_\sigma + \frac{g^2_{\omega NN}}{4\pi \cdot 10.6} T_\omega + \frac{g^2_{\rho NN}}{4\pi \cdot 0.43} T_\rho)$$

$$+ \frac{g_{\rho NN} g_{\omega NN}}{4\pi \sqrt{0.43 \cdot 10.6}} \frac{\langle \omega | H | \rho^0 \rangle}{(-4300 \text{ MeV}^2)} T_{\rho\omega} + T_{WF}, \qquad (38)$$

where

$$\tilde{\beta} = \frac{g_{\eta NN}}{\sqrt{4\pi \cdot 0.10}} \frac{\langle \eta | H | \pi^0 \rangle}{-4200 \text{ MeV}^2}, \qquad (39)$$

and

$$\begin{array}{llll}
T_\pi & = & -0.76 + 0.74i & \quad T_\omega & = & -0.15 + 0.15i \\
\bar{T}_\pi & = & -0.14 + 0.14i & \quad T_\rho & = & -0.08 + 0.08i \\
T_{OB} & = & -0.86 + 0.71i & \quad T_{\rho\omega} & = & -0.84 + 1.1i \\
T_\sigma & = & -0.14 + 0.19i & \quad T_{WF} & = & 0.41 - 0.14i.
\end{array} \qquad (40)$$

The units of the terms $T$ of Eq. (38) are $10^{-4}$ fm$^{-2}$, and the numerical values are presented in terms of these units. $T_\pi$ and $\bar{T}_\pi$ are the seagull terms from Fig. 5a; $T_{OB}$ is the one-body term from Fig. 5b; the Z-graph terms are $T_\sigma$, $T_\omega$, and $T_\rho$; the $\rho^0 - \omega$ Z-graph term is $T_{\rho\omega}$; and CSB in the final state wavefunction enters through $T_{WF}$. We see that the leading order term now is of leading numerical order. The specific values of $T_\pi$ and $\bar{T}_\pi$ vary considerably with the chosen internucleon force (141). Using these results, the cross section at 228.5 MeV is:

$$\sigma = 4.3 \left| \frac{T}{10^{-4} \text{ fm}^{-2}} \right|^2 \text{ pb} = 64 \text{ pb} \qquad (41)$$

If the Z-graphs are eliminated the cross section would be 15 pb, so that (at least) the leading terms are in agreement with the experiment. Another possibility is that relative signs of the $\eta-$ and $\pi-$ nucleon coupling constants differ. In this case, the computed cross section obtained from using the range of Eq. (9) would vary between 1.6 and 10.5 pb. The present results represent a reasonable starting point for future computations, but the Z-graphs and the sign of $g_{\eta NN}$ need to be better understood. Given the dramatic influence of the ISI in such calculations, it is of paramount importance that new experimental constraints be obtained from deuteron-deuteron interactions. From a theoretical standpoint, the necessary task of constructing the initial and final wave functions and the pion production operator from a single consistent EFT remains to be done.

## 6   Summary

Charge symmetry breaking in the strong interaction occurs because of the difference between the mass of the up and down quarks. CSB is an intrinsically quark effect, and the use of EFT allows us to follow the influence of confined quarks on phenomena in hadrons and nuclei.

The last ten years have seen very significant developements in the understand-ing of CSB. The use of an EFT for QCD (Sect. 3) has led to a reasonable fundamental understanding of the strong NN interaction at all but the shortest of ranges. EFT has also allowed a similar fundamental understanding of the different classes of charge independence breaking and CSB interactions (Sect. 3.3). Significant problems such as understanding the isospin breaking of $\pi - N$ scattering (Sect. 3.3.2) and CSB in hypernuclei (Sect. 2.5.3) remain to be solved.

An especially significant prediction of EFT is that CSB is very important in the interactions between neutral pions and nucleons, as represented by a particular seagull term. This observation led to the suggestion that experiments measuring the CSB production of a neutral pion would shed essential light on this important prediction (Sect. 3.3). The challenge was taken up with the ultimately successful measurements of CSB in the forward-backward asymmetry of the $np \to d\pi^0$ reaction (Sect. 4) and in the non-zero cross section for the $dd \to \alpha\pi^0$ reaction (Sect. 5).

Some interesting first steps in understanding the sizes of the recently observed CSB effects have been made. Power counting techniques have been developed recently for the generally difficult task of understanding strong pion production in few nucleon collisons (Sect. 3.4). It is necessary to use the same EFT to construct the pion production operators and the initial state and final state wave functions. It now seems that the necessary tools are available.

Understanding CSB pion production presents its own difficulties as many terms contribute. Realistic two and three nucleon forces are used to construct initial and final state wave functions needed to compute CSB $\pi^0$ production observables (Sect. 5.2) and using these wave functions has profound influences on the calculation. The CSB seagull terms are responsible for the positive sign of the forward-backward asymmetry in the $np \to d\pi^0$ reaction (Sect. 4.2). The CSB seagull terms are important in the $dd \to \alpha\pi^0$ reaction, once the effects of initial state interactions are included. Much remains to be done before any precise statements can be made. But optimism is well-founded. We are at the threshold of understanding how the light-quark mass difference makes its presence felt in nuclear physics.

## Acknowledgements

This work was supported by U.S. Department of Energy under grant DE-FG02-97ER41014 and by U.S. National Science Foundation under grants PHY-98-02872, PHY-01-00348, PHY-02-44999, and PHY-04-57219. We also thank A.D. Bacher, S. Gardner, K. Maltman, A. Schwenk, A.W. Thomas, and U. van Kolck for useful discussions and K. Maltman for providing unpublished results. We thank the CSB theory collaboration: A.C. Fonseca, A. Gårdestig, C. Hanhart, C.J. Horowitz, G.A. Miller, J.A. Niskanen, A. Nogga, and U. van Kolck for sharing their insights and results prior to publication.

# References

1. Miller GA, Nefkens BMK, Slaus I. *Phys. Rept.* 194:1 (1990)
2. Stephenson EJ *et al. Phys. Rev. Lett.* 91:142302 (2003)
3. Opper AK *et al. Phys. Rev. Lett.* 91:212302 (2003)
4. Hughes EW, Voss R. *Ann. Rev. Nucl. Part. Sci.* 49:303 (1999)
5. Kaplan DB, Manohar A. *Nucl. Phys.* B310:527 (1988)
6. McHeown RD. *Phys. Lett.* B219:140 (1989); Beck DB. *Phys. Rev.* D39:3248 (1989)
7. Aniol KA *et al. Phys. Lett.* B509:211 (2001), *Phys. Rev.* C69:065501 (2004), [arXiv:nucl-ex/0506011]
8. Armstrong DS *et al. Phys. Rev. Lett.* 95:092001 (2005)
9. Miller GA. *Phys. Rev.* C57:1492 (1998)
10. Zeller GP *et al. Phys. Rev. Lett.* 88:091802 (2002); Erratum-ibid. 90:239902 (2003)
11. Paschos EA, Wolfenstein L. *Phys. Rev.* D7:91 (1973)
12. Sather E. *Phys. Lett.* B274:433 (1992)
13. Rodionov EN, Thomas AW, Londergan JT. *Mod. Phys. Lett.* A9:1799 (1994)
14. Londergan JT, Thomas AW. *Phys. Lett.* B558:132 (2003), *Phys. Rev.* D67:111901 (2003)
15. Bennett GW *et al. Phys. Rev. Lett.* 92:161802 (2004)
16. Hertzog DW, Morse WM. *Ann. Rev. Nucl. Part. Sci.* 54:141 (2004)
17. Davier M, Marciano WJ. *Ann. Rev. Nucl. Part. Sci.* 54:115 (2004)
18. Gourdin M, DeRafael E. *Nucl. Phys.* B10:667 (1969)
19. Maltman K, Wolfe CE. [arXiv:hep-ph/0509224]; Maltman K, private communication (2006)
20. Eidelman S *et al.* (Particle Data Group). *Phys. Lett.* B592:1 (2004); 2005 partial update for 2006 edition; [URL http://pdg.lbl.gov]
21. Quenzer A *et al. Phys. Lett.* B76:512 (1978); Barkov LM *et al. Nucl. Phys.* B256:365 (1985)
22. Gardner S, O'Connell HB. *Phys. Rev.* D57:2716 (1998), Erratum-ibid. D62:019903 (2000)
23. Coon SA, Barrett RC. *Phys. Rev.* C36:2189 (1987)
24. Akhnetshin RR *et al.* [CMD-2]. *Phys. Lett.* B578:285 (2004); Aloisio A *et al.* [KLOE]. *Phys. Lett.* B606:12 (2005); Achasov MN *et al.* arXiv:hep-ex/0506076
25. Langacker P. *Phys. Rev.* D20:2983 (1979)
26. Maltman K, O'Connell HB, Williams AG. *Phys. Lett.* B376:19 (1996)
27. Piekarewicz J, Williams AG *Phys. Rev. C* 47:2462 (1993)
28. O'Connell *et al. Prog. Part. Nucl. Phys.* 39:201 (1997)
29. Miller GA. *Chin. J. Phys.* 32:1075 (1994) [arXive:nucl-th/9406023]
30. Miller GA, van Oers WTH, in *Symmetries and fundamental interactions in nuclei* ed. by Haxton WC and Henley EM (1995) World Scientific [arXiv:nucl-th/9409013]
31. Bauer TH, Spital RD, Yennie DR, Pipkin FM. *Rev. Mod. Phys.* 50:261 (1978)
32. O'Connell HB *et al. Phys. Lett.* B370:12 (1996)

33. van Kolck U, Friar JL, Goldman T. *Phys. Lett.* B371:169 (1996)

34. Coon SA, Scandron MD. *Phys. Rev.* C51:2923 (1995)

35. Cohen TD, Miller GA *Phys. Rev.* C 52:3428 (1995)

36. Dumbrais O. *Nucl. Phys.* B216:277 (1983)

37. Reuber A, Holinde K, Speth J. *Nucl. Phys.* A570:541 (1991); Machleidt R. *Phys. Rev.* C63:024001 (2001)

38. Grein W and Kroll P. *Nucl. Phys.* A338:332 (1980); *Nucl. Phys.* A377:505 (1982)

39. Deans SR Wooten JW. *Phys. Rev.* 185:1797 (1969)

40. Tiator L *et al.* *Nucl. Phys.* A580:455 (1994)

41. Gårdestig A. *et al.* *Phys. Rev.* C69:044606 (2004)

42. van Kolck U *et al.* *Phys. Rev. Lett.* 80:4386 (1998)

43. Nielsen M. hep-ph/0510277; Kroll P. *Mod. Phys. Lett.* A20:2667 (2005)

44. Henley EM, Miller GA. in *Mesons in Nuclei* edited by Rho M and Wilkerson D (1979) North-Holland

45. Cheung CY, Henley EM, Miller GA. *Phys. Rev. Lett.* 43:1215 (1979); *Nucl. Phys.* A305:342 (1978); A348:365 (1980)

46. Ericson TEO, Miller GA. *Phys. Lett.* B132:32 (1983)

47. Henley EM. in "Isospin In Nuclear Physics" ed. by Wilkerson DH (North-Holland, Amsterdam 1969) 17

48. Gabioud B *et al.* *Nucl. Phys.* A420:496 (1984); de Téramond GF, Gabioud B. *Phys. Rev.* C36:691 (1987)

49. Schori O *et al.* *Phys. Rev.* C35:2252 (1987)

50. Howell CR. *Phys. Lett.* B444:252 (1998)

51. Gibbs WR, Gibson BF, Stephenson GJ Jr. *Phys. Rev.* C11:90 (1975); 12:2130(E) (1975); 16:327 (1977); 17:856(E) (1978)

52. Gårdestig A, Phillips DR. *Phys. Rev.* C73:014002 (2006)

53. Huhn V *et al.* *Phys. Rev. Lett.* 85:1190 (2000).

54. González Trotter DE *et al.* *Phys. Rev. Lett.* 83:3788 (1999)

55. Gersten A. *Phys. Rev.* C18:2252 (1978)

56. Abegg R *et al.* *Phys. Rev. Lett.* 56:2571 (1986); *Phys. Rev.* D39:2464 (1989)

57. Abegg R *et al.* *Phys. Rev.* C57:2126 (1998)

58. Knutson LD *et al.* *Phys. Rev. Lett.* 66:1410 (1991); Vigdor SE *et al.* *Phys. Rev.* C46:410 (1992)

59. Miller GA, Thomas AW, Williams AG. *Phys. Rev. Lett.* 56:2567 (1986); Williams AG, Thomas AW, Miller GA. *Phys. Rev.* C36:1956 (1987)

60. Holzenkamp B, Holinde K, Thomas AW. *Phys. Lett.* B195:121 (1987)

61. Iqbal MJ, Niskanen JA. *Phys. Rev.* C38:2259 (1988)

62. Okamoto K. *Phys. Lett.* 11:150 (1964)

63. Friar JL. *Nucl. Phys.* A156:43 (1970)

64. Brandenburg RA, Coon SA, Sauer PU. *Nucl. Phys.* 294:305 (1978); Wu Y, Ishikawa S, Sasakawa T. *Phys. Rev. Lett.* 64:1875 (1990)

65. Machleidt R, Muther H. *Phys. Rev.* C63:034005 (2001)

66. Coon SA, Niskanen JA. *Phys. Rev.* C53:1155 (1996)
67. Nolen JA, Schiffer JP. *Ann. Rev. Nucl. Part. Sci.* 19:471 (1969)
68. Negele JW. *Nucl. Phys.* A165:305 (1971)
69. Shlomo A. *Rep. Prog. Phys.* 41:957 (1978)
70. Blunden PG, Iqbal MJ. *Phys. Lett.* B198:14 (1987)
71. Zuker AP et al. *Phys. Rev. Lett.* 89:142502 (2002)
72. Gal A. *Adv. Nucl. Phys.* 8:1 (1975)
73. Gibson BF. *Nucl. Phys.* A479:115c (1986)
74. Nogga A, Kamada H, Glöckle W. *Phys. Rev. Lett.* 88:172501 (2002); Nogga A/ *Nucl. Phys.* A754:36 (2005)
75. Weinberg S. *Physica* A96:327 (1979)
76. van Kolck U. *Prog. Part. Nucl. Phys.* 43:337 (1999)
77. Bedaque PF, van Kolck U. *Ann. Rev. Nucl. Part. Sci.* 52:339 (2002)
78. Hanhart C. *Phys. Rept.* 397:155 (2004)
79. Goldstone J, Salam A, Weinberg S. *Phys. Rev.* 127:965 (1962)
80. Bernard V, Kaiser N, Meißner U-G. *Int. J. Mod. Phys.* E4:193 (1995)
81. van Kolck U. *Prog. Part. Nucl. Phys.* 43:337 (1999)
82. Weinberg S. *Phys. Lett.* B251:288 (1990); *Nucl. Phys.* B363:3 (1991)
83. Coon SA, Friar JL. *Phys. Rev.* C34:1060 (1986); Friar JL. *Czech. J. Phys.* 43:259 (1993)
84. Kaplan DB, Savage MJ, Wise MB. *Phys. Lett.* B424:390 (1998); *Nucl. Phys.* B534:329 (1998)
85. Beane SR et al. *Nucl. Phys.* A700:377 (2002); Nogga A, Timmermans RGE, van Kolck U. *Phys. Rev.* C72:054006 (2005)
86. Ordóñez C, van Kolck U. *Phys. Lett.* B291:459 (1992)
87. Ordóñez C, Ray L, van Kolck U. *Phys. Rev. Lett.* 72:1982 (1994); *Phys. Rev.* C53:2086 (1996)
88. Wiringa RB, Stoks VGJ, Schiavilla R. *Phys. Rev.* C51:38 (1995)
89. Kaiser N, Brockmann R, Weise W. *Nucl. Phys.* A625:758 (1997); Kaiser N, Gerstendörfer S, Weise W. *Nucl. Phys.* A637:395 (1998); Ballot JL, Robilotta MR, da Rocha CA. *Phys. Rev.* C57:1574 (1998)
90. Rentmeester MCM et al. *Phys. Rev. Lett.* 82:4992 (1999); Rentmeester MCM, Timmermans RGE, de Swart JJ. *Phys. Rev.* C67:044001 (2003)
91. Entem DR, Machleidt R. *Phys. Lett.* B524:93 (2002)
92. Entem DR, Machleidt R. *Phys. Rev.* C68:041001 (2003)
93. Epelbaum E, Glöckle W, Meißner U-G. *Nucl. Phys.* A747:362 (2005)
94. van Kolck U. *Few-Body Syst. Suppl.* 9:444 (1995); U. of Texas Ph.D. Dissertation (1993)
95. Weinberg S. *Trans. New York Acad. Sci.* 38:185 (1977)
96. van Kolck U, Niskanen JA, Miller GA. *Phys. Lett.* B493:65 (2000)
97. Gasser J, Leutwyler H. *Phys. Rep.* 87:77 (1982)
98. Friar JL, van Kolck U. *Phys. Rev.* C60:034006 (1999)
99. Coon SA, McKellar BHJ, Stoks VGJ. *Phys. Lett.* B385:25 (1996)
100. Niskanen JA. *Phys. Rev.* C65:037001 (2002)

101. Friar, JF *et al.* *Phys. Rev.* C68:024003 (2003)
102. Li GQ, Machleidt R. *Phys. Rev.* C58:1393 (1998)
103. Li GQ, Machleidt R. *Phys. Rev.* C58:3153 (1998)
104. Walzl M, Meißner U-G, Epelbaum E. *Nucl. Phys.* A693:663 (2001)
105. Epelbaum E, Meißner U-G. *Phys. Lett.* B461:287 (1999)
106. Valcarce A *et al.* *Rept. Prog. Phys.* 68:965 (2005)
107. Enter DR, Fernandez F, Valcarce A. *Phys. Lett.* 463:153 (1999)
108. Nasu T, Oka M, Takeuchi S. *Phys. Rev.* C68:024006 (2003); C69:029903(E) (2004)
109. Miller GA. *Phys. Rev.* C39:1563 (1989)
110. Gibbs WR, Ai L, Kaufman WB. *Phys. Rev. Lett.* 74:3740 (1995); Matsinos E. *Phys. Rev.* C56:3014 (1997)
111. Fettes N, Meißner U-G, Steininger S. *Phys. Lett.* B451:233 (1999)
112. Fettes N, Meißner U-G. *Phys. Rev.* C63:045201 (2001)
113. Cohen TD *et al.* *Phys. Rev.* C53:2661 (1996)
114. Hanhart C, van Kolck U, Miller GA. *Phys. Rev. Lett.*, 85:2905 (2000)
115. Hanhart C, Kaiser N. *Phys. Rev.* C66:054005 (2002)
116. Bernard V, Kaiser N, Meißner U-G. *Eur. Phys. J.* A4:259 (1999)
117. A. Gardestig, D. R. Phillips and C. Elster, arXiv:nucl-th/0511042, *Phys. Rev.* CXX:XYZT (2006)
118. Lensky V *et al.* arXiv:nucl-th/0511054
119. Bartlett DF *et al.* *Phys. Rev.* D1:1984 (1970)
120. Wilson SS *et al.* *Nucl. Phys.* B33:253 (1971)
121. Hollas CL *et al.* *Phys. Rev.* C24:1561 (1981)
122. Hutcheon DA *et al.* *Nucl. Phys.* A535:618 (1991)
123. Walden PL *et al.* *Nucl. Inst. Meth.* A421:142 (1999)
124. Helmer R *et al.* *Can. J. Phys.* 65:588 (1987)
125. Hutcheon DA *et al.* *Nucl. Inst. Meth.* A459:448 (2001)
126. Auce A *et al.* *Phys. Rev.* C53:2919 (1987); Okamura H *et al.* *Phys. Rev.* C58:2180 (1998); Baumer C *et al.* *Phys. Rev.* C63:037601 (2001)
127. Press WH *et al.* *Numerical Recipes* (1992)
128. Niskanen JA, Sepestyen M, Thomas AW. *Phys. Rev.* C38:838 (1988); Niskanen JA. *private communication*
129. Niskanen JA, Iqbal MJ. *Phys. Lett.* B218:272 (1989)
130. Niskanen JA. *Few-Body Systems* 26:241 (1999)
131. Coon SA *et al.* *Phys. Rev.* D34:2784 (1986)
132. www.physics.arizona.edu/~vankolck/coolerCSBtheory.html
133. Pickar MA *et al.* *Phys. Rev.* C46:397 (1992)
134. Greider KR. *Phys. Rev.* 122:1919 (1961)
135. Banaigs J *et al.* *Phys. Rev. Lett.* 58:1922 (1987)
136. Lapidus LI. *Zh. Eksp. Teor. Fiz.* 31:865 (1956); *Sov. Phys. JETP* 4:740 (1957)
137. Goldzahl L *et al.* *Nucl. Phys.* A533:657 (1991)
138. Doborkhotov D, Fäldt G, Gårdestig A, Wilkin C. *Phys. Rev. Lett.* 83:5246 (1999)
139. Ermisch K *et al.* *Phys. Rev.* C71:064004 (2005)

140. Sekiguchi K *et al. Phys. Rev. Lett.* 95:162301 (2005)
141. Nogga A *et al.* [CSB Theory Collaboration]. nucl-th/0602003
142. Friar JL Huber D van Kolck U. *Phys. Rev.* C59:53 (1999) Coon SA, Han HK. *Few Body Syst.* 30:131 (2001)